%% file: main.tex
\documentclass[iop]{emulateapj}
\usepackage{bibStyle/apjfonts}
\usepackage{graphicx}

\usepackage{color}
\usepackage{epstopdf}
\usepackage{amsmath}
\usepackage{float}

\def\ltsima{$\; \buildrel < \over \sim \;$}
\def\simlt{\lower.5ex\hbox{\ltsima}}
\def\gtsima{$\; \buildrel > \over \sim \;$}
\def\simgt{\lower.5ex\hbox{\gtsima}}

\newcommand\lsim{\mathrel{\rlap{\lower4pt\hbox{\hskip1pt$\sim$}}
\raise1pt\hbox{$<$}}}
\newcommand\gsim{\mathrel{\rlap{\lower4pt\hbox{\hskip1pt$\sim$}}
\raise1pt\hbox{$>$}}}

\usepackage{color}

 \newcommand{\tot}{\mathrm{tot}}
\shorttitle{SNe Kicks in hierarchical triple systems}
\shortauthors{C. X. Lu \& S. Naoz}

\begin{document}

\title[SNe Kicks in hierarchical triples]{Supernovae Kicks in hierarchical triple systems}

\author{Cicero X. Lu$^{1,2}$  \&  Smadar Naoz$^{1,3}$ }

\altaffiltext{1}{Department of Physics and Astronomy, University of California, Los Angeles, CA 90095, USA}
\altaffiltext{2}{Department of Physics and Astronomy, Johns Hopkins University, MD 21218, USA}
\altaffiltext{3}{Mani L. Bhaumik Institute for Theoretical Physics, Department of Physics and Astronomy, UCLA, Los Angeles, CA 90095, USA}
\email{cicerolu@jhu.edu \\ snaoz@astro.ucla.edu}

\begin{abstract}
Most massive stars, if not all, are in binary configuration or higher multiples. These massive stars undergo supernova explosions and end their lives as either black holes or neutron stars. Recent observations have suggested that neutron stars and perhaps even black holes receive large velocity kicks at birth. Such natal kicks and the sudden mass loss can significantly alter the orbital configuration of the system. Here we derive general analytical expressions that describe the effects of natal kicks in binaries on hierarchical triple systems. We explore several proof-of-concept applications such as black hole and neutron stars binaries and X-ray binaries with either stellar or Supermassive Black Hole (SMBH) companions on a wide orbit. Kicks can disrupt the hierarchical configuration, although it is harder to escape the potential well of an SMBH. Some binary systems do escape the SMBH system resulting in hyper-velocity binary system. Furthermore, kicks can result in increasing or decreasing the orbital separations. Decreasing the orbital separation may have significant consequences in these astrophysical systems. For example, shrinking the separation post-supernova kick can lead to the shrinking of an inner compact binary that then may merge via Gravitational Wave (GW) emission. This process yields a supernova that is shortly followed by a possible GW-LIGO event. Interestingly, we find that the natal kick can result in shrinking the outer orbit, and the binary may cross the tertiary Roche limit, breaking up the inner binary. Thus, in the case of SMBH companion, this process can lead to either a tidal disruption event or a GW-LISA detection event (Extreme Mass ratio inspiral, EMRI)   with a supernova precursor.
\end{abstract}

\keywords{stars: binaries: close, stars: black holes, evolution, kinematics and dynamics, X-rays: binaries}

\maketitle

\input{chapters/introduction}
\input{chapters/system_setup}
\input{chapters/post_SN_orbital_param}

\input{chapters/applications}
\input{chapters/GW_source/ns_bh}
\input{chapters/GW_source/bhb}
\input{chapters/GW_source/ns_ns}

\input{chapters/Xray_binary/ns_lmxb}
\input{chapters/Xray_binary/bh_lmxb}

\input{chapters/discussion}
\acknowledgements
We thank the referee for a thoughtful read of the paper and comments that helped strengthen our paper. We thank Alexander Stephan and Bao-Minh Hoang  for useful discussion and suggestions. 
CL acknowledges the UCLA Van Tree Foundation, Undergraduate Research Scholars Program Scholarship as well as the Stone Foundation Summer Research Scholarship. CL thanks Izuki Matsuba for discussion and help. SN acknowledges the partial support of the Sloan fellowship and from the NSF through grant No. AST-1739160.

\bibliographystyle{bibStyle/hapj}
\bibliography{bibStyle/Kozai}
\appendix
\section{A. The post-SN angular momentum}\label{App:J1}
Opening Equation (\ref{eq:e1n})
\begin{eqnarray}\label{eq:J1n}
J_1^2&=&\frac{1 + u_k^2 - 2 \beta + 2 u_k \cos\theta + 
 e_1 \cos E_1  (1 + u_k^2 + 2 u_k \cos \theta ) }{\beta(e_1\cos E_1 -1)} \bigg( 1 + u_k^2 - e_1^2 (1 + u_k^2) \cos^2E_1
  - e_1^2 \sin^2 E_1 + \frac{1}{2} u_k (-1 + e_1 \cos E_1 ) \nonumber \\ 
  &\times& \bigg[ 2 u_k (1 + e_1 \cos E_1) \cos^2 \alpha -  4 (1 + e_1 \cos E_1) \cos \theta + 4 e_1 \sqrt{\frac{1 +e_1\cos E_1}{e_1 
  \cos E_1 -1   }} \cos\alpha \sin E_1  \bigg] \bigg)  
\end{eqnarray}

\section{B. The tilt angle}\label{App:tilt}

The tilt angle is defined as the angle between the plane of the post-SN orbital plane and the pre-SN plane. \citet{Kalogera00} studied the tilt angle following a SN-kick in a circular binary. The tilt angle can be related to spin-orbit misalignment which can affect their resulted gravitational radiation waveforms of  coalescing compact binaries, and thus affect their detectability. 
Using the eccentricity vector:
\begin{equation}
{\bf e}=\frac{1}{\mu}\left(\dot{\bf r}\times{\bf h}-\mu\frac{{\bf r}}{r}\right)  \ .
\end{equation}
where $\mu=G(m_1+m_2)$, we can find the tilt angle between the post and pre SN explosion, which we denote as $\Delta \psi$.
This angle is simply:
\begin{equation}\label{eq:psi}
\cos\Delta\psi_1=\frac{{\bf e}_{1,n} \cdot {\bf e}_1}{e_{1,n} e_1}  \ ,
\end{equation}
where $e_{1,n}$ is found using Equation (\ref{eq:e1n}). 

\section{C. Post and pre-SN velocity relations} 
The velocity vector to the inner orbit which is associated with ${\bf r}={\bf r}_2-{\bf r}_1$ (see Figure  \ref{fig:Cartoon}) is defined by ${\bf v}_r={\bf v}_2-{\bf v}_1$.
The velocity vector of the outer orbit is defined by ${\bf V}_3={\bf v}_3-{\bf v}_{\rm c.m.}$, where ${\bf v}_3$ is the velocity vector associated with the position vector ${\bf r}_3$ and ${\bf v}_{\rm c.m.}$ is the velocity vector of the inner orbit center of mass associated with the center of mass position vector ${\bf r}_{\rm c.m.}$ (see Figure \ref{fig:Cartoon}).  Note that:
\begin{equation}\label{eq:vcm}
{\bf v}_{\rm c.m.}=\frac{m_1{\bf v}_1+m_2{\bf v}_2}{m_1+m_2} \ .
\end{equation}
As $m_2$ undergoes SNe, we find that  ${\bf v}_{r,n}={\bf v}_2 + {\bf v}_k-{\bf v}_1 = {\bf v}_r + {\bf v}_k$ and thus the new outer orbit velocity is
\begin{equation}\label{eq:vR3}
{\bf V}_{3,n}={\bf V}_3-\frac{m_1(m_{2,n}-m_2) {\bf v}_r}{(m_1+m_{2,n})(m_1+m_2)} + \frac{m_{2,n}}{m_1+m_{2,n} }{\bf v}_k  \quad {\rm for} \quad m_2\to m_{2,n} \ .
\end{equation}
Note that if $m_1$ undergoes SN then:
\begin{equation}\label{eq:vR3m1}
{\bf V}_{3,n}={\bf V}_3-\frac{m_2(m_{1}-m_{1,n}) {\bf v}_r}{(m_2+m_{1,n})(m_1+m_2)} - \frac{m_{1,n}}{m_2+m_{1,n} }{\bf v}_k  \quad {\rm for} \quad  m_1\to m_{1,n} \ .
\end{equation}
Note that all the vectors need to be rotated with respect to the invariable plane.

\section{D. Orbital parameter plots for the BH-BH system}\label{App:Plots} 

\begin{figure*}[t!]
\begin{center}
\includegraphics[width=\linewidth, angle=90, scale = 1.35]{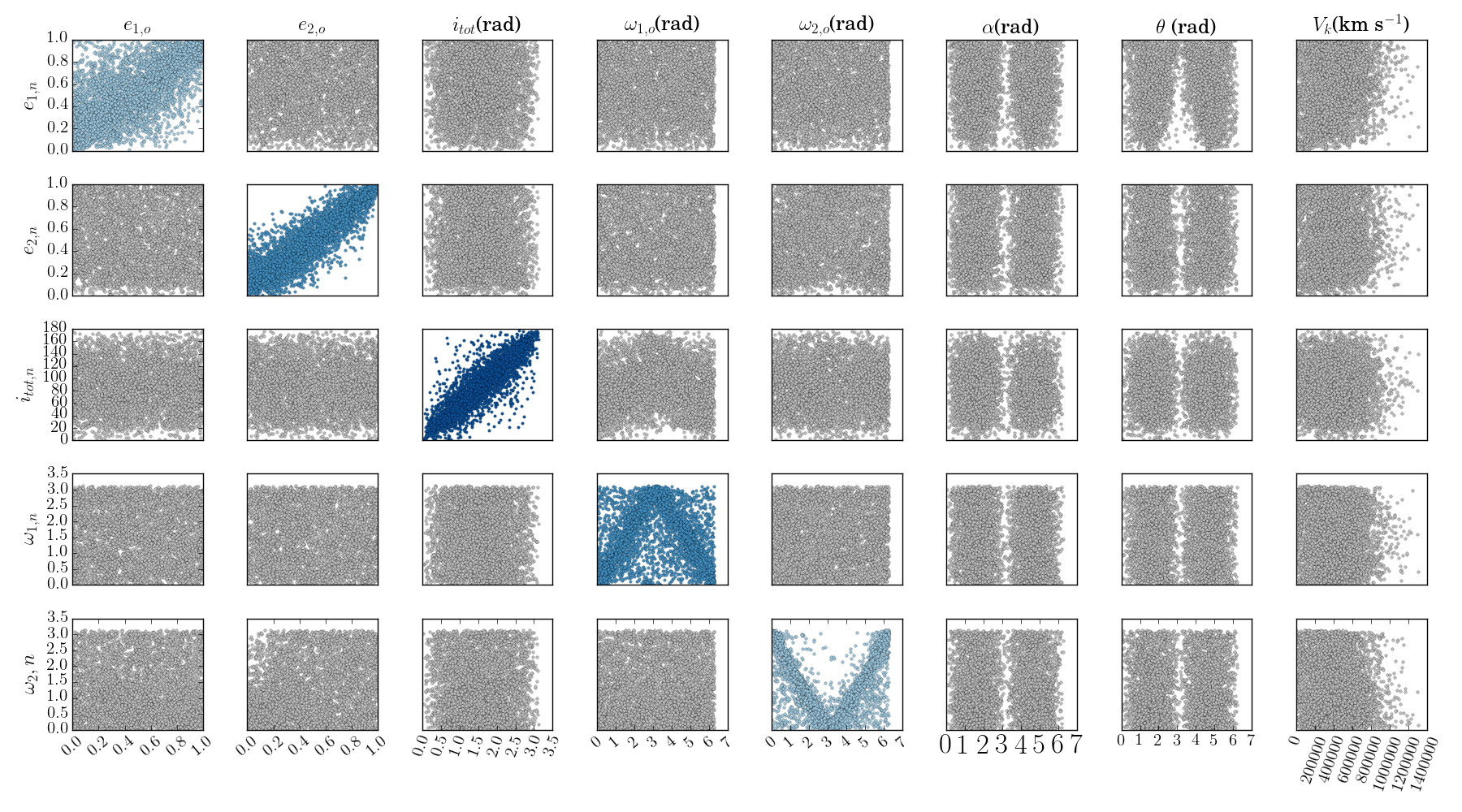}
\caption{\upshape BHB system(with SMBH tertiary) after first SN and its resulting changes in parameters. This is for systems with $a_1$ = ~$5 R_\odot$ and $a_2$ = ~$1000$AU. The subscript "o" means pre-SN orbital parameter values and the subscript "n" stands for the values of orbital parameters after first SN.}\label{fig:bigplot_1sn}
\end{center}
\end{figure*}

\begin{figure*}[t!]
\begin{center}
\includegraphics[width=\linewidth, angle=90, scale = 1.35]{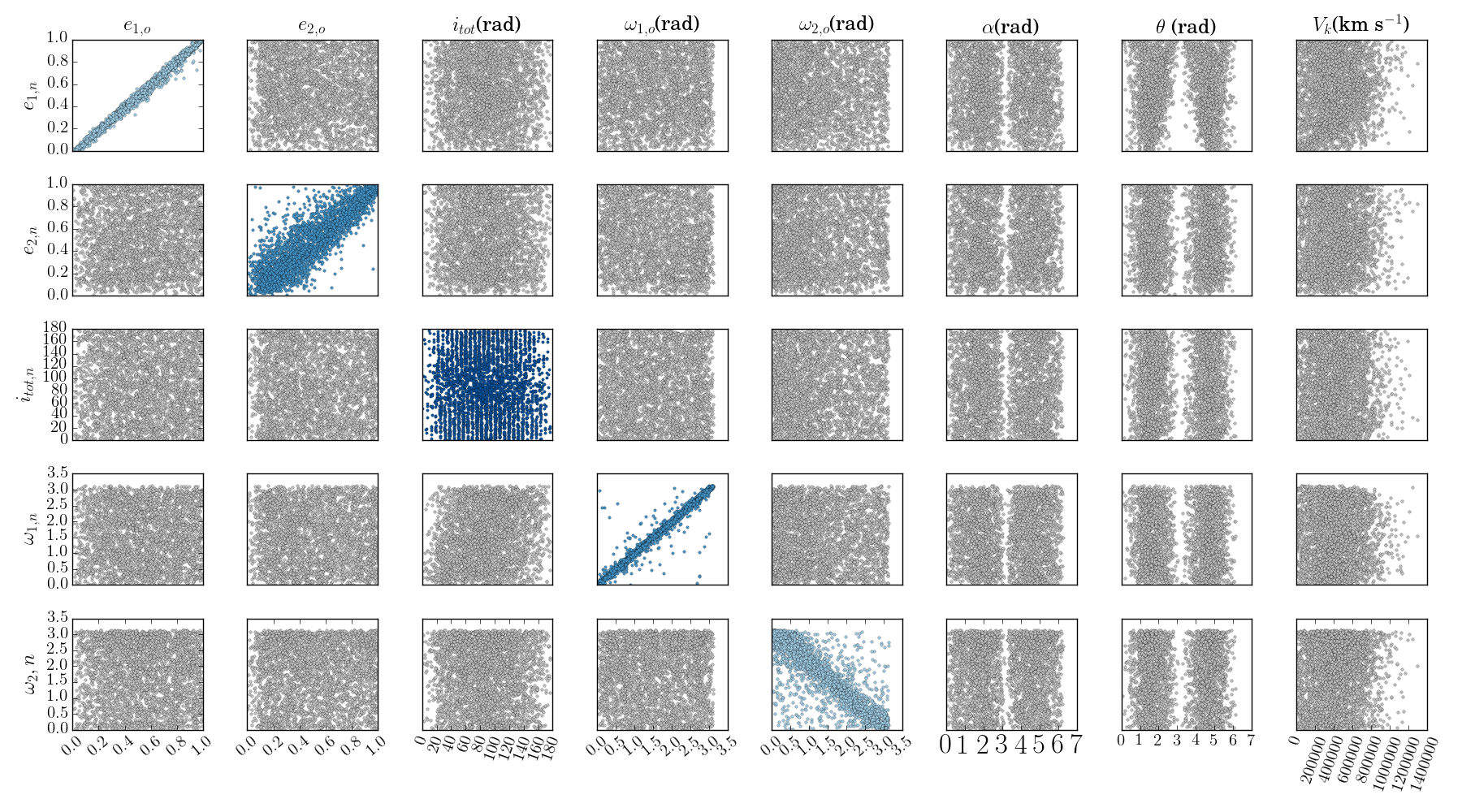}
\caption{\upshape BHB (with SMBH tertiary) system after 2nd SN and its resulting changes in parameters. This is for systems with $a_1$ = ~$5 R_\odot$ and $a_2$ = ~$1000$AU. The subscript "o" means post-1st SN orbital parameter values and the subscript "n" stands for the values of orbital parameters after second SN.}\label{fig:bigplot_2sn}
\end{center}
\end{figure*}

Motivated by the recent LIGO detection we provide the orbital parameters distribution of one of our proof-of-concept runs. Specifically, we chose to show the case for which $a_1=5$~$R_{\odot}$, $a_2=1000$~AU and a SMBH tertiary \citep[see][]{Hoang+17}. This system is shown in Figures \ref{fig:bigplot_1sn} and \ref{fig:bigplot_2sn}. 

\section{E. Runs with wider inner binary }\label{App:5au}
 Here we present a Table with the Monte-Carlo results while setting $a_1=5$~au. To allow comparison we reiterate the nominal results considered throughout the paper. All these runs are note as $A5$, where $A$ is the nominal runs presented in Table \ref{table}.

%================================================================================
%                               An alternative table with A1=5 AU
%================================================================================
\begin{table*}
\centering
 \begin{tabular}{l |l || c c c c c c c || c | c | c | c | c} 
 \hline
 Name & Sim &$m_1$ & $m_{1,n}$ & $m_2$    & $m_{2,n}$ & $m_3$     & $a_1$ & $a_2$ &\% Bin & \%  Triples & \% in $R_{\rm Roche,in}$   & \% in $R_{\rm Roche,out}$ & \%escaped \\ 
& & M$_\odot$& M$_\odot$&M$_\odot$ &M$_\odot$  & M$_\odot$ & R$_\odot$    & AU &  out of total & out of Bin &out of Bin & out of 3 &  Bin \\ 
 \hline\hline

  NS-LMXB & (a) & 4     & 1.4  & 1     & 1     & 3     & MC$^1$     & MC$^{2,EC}$ &  4   & 0 & 0 & 0 &  100\\
  & (b) & 4     & 1.4  & 1     & 1     & $4\times 10^6$     & MC$^1$     & MC$^{2,EC}$  &   4 &  94 & 0 & 4  & 6 \\
  & (c) & 4     & 1.4  & 1     & 1     & $4\times 10^6$     & MC$^1$     & MC$^{2,BW}$  & 4  & 99  & 0 &   2 & 1\\
   &(a5)& 4     & 1.4  & 1     & 1     & 3     & 1075.5$^\dagger$    & MC$^{2,EC}$ &  0.2  & 19 & 0 & 25 &  0\\
  &(b5) & 4 & 1.4  & 1     & 1     & $4\times 10^6$     & 1075.5$^\dagger$     & MC$^{2,EC}$  &   1 &  100 & 0 & 33  & 0\\
  &(c5) & 4     & 1.4  & 1     & 1     & $4\times 10^6$     & 1075.5$^\dagger$     & MC$^{2,BW}$  & 0.3  & 100 &0 & 15 & 0\\
  \hline

 BH-LMXB & (d) & 9     & 7    & 1     & 1     & 3     & MC$^1$    & MC$^{2,EC}$ &   11   & 1 & 24 & 13   & 99\\
 & (e) & 9     & 7    & 1     & 1     &  $4\times 10^6$      & MC$^1$  &   MC$^{2,EC}$ &   11   &  99 & 24 & 7 & 1\\
 & (f) & 9     & 7    & 1     & 1     &  $4\times 10^6$      & MC$^1$  &   MC$^{2,BW}$ &  10   &  92  & 25   & 2 & 8 \\
  &(d5) & 9     & 7    & 1     & 1     & 3           & 1075.5$^\dagger$    & MC$^{2,EC}$ &   1   & 15 & 0 & 13   & 0.1\\
 & (e5) & 9     & 7    & 1     & 1     &  $4\times 10^6$    & 1075.5$^\dagger$   &   MC$^{2,EC}$ &   1   &  100 & 2 & 17 & 0\\
 & (f5)& 9     & 7    & 1     & 1     &  $4\times 10^6$     & 1075.5$^\dagger$   &   MC$^{2,BW}$ &  1  &  100  & 2   & 14 & 0 \\
  \hline

  NS-BH & (g)  & 4     & 1.4  & 10    & 10    & 3     & 5    & 1000 & 33 & 0  & 0 & 0 & 100\\
  & (h) & 4     & 1.4  & 10    & 10    & 3     & 5    & MC$^{2,EC}$ & 33 & 0  & 0 & 0 & 100\\
  & (i) & 4     & 1.4  & 10    & 10    & $4\times 10^6$     & 5    & MC$^{2,BW}$ & 33  & 99  & 0 & 0 & 1\\
 & (j) & 4     & 1.4  & 10    & 10    & $4\times 10^6$     & MC$^1$  & MC$^{2,EC}$ & 12  & 71  & 0 & 2 & 29\\
  & (k)& 4     & 1.4  & 10    & 10    &  $4\times 10^6$     & 5    &   1000 & 33& 100 & 0 & 1 & 0\\
  & (g5)& 4     & 1.4  & 10    & 10    & 3     & 1075.5$^\dagger$    & 1000 & 2 & 22.4  & 0 & 3 & 0.1\\
 & (h5)&4     & 1.4  & 10    & 10    & 3     & 1075.5$^\dagger$    & MC$^{2,EC}$ & 2 & 29  & 0 &11 & 0.2\\
& (i5)& 4     & 1.4  & 10    & 10    & $4\times 10^6$     & 1075.5$^\dagger$     & MC$^{2,BW}$ & 2  & 100  & 0 & 7 & 0 \\
 &(j5)& 4     & 1.4  & 10    & 10    & $4\times 10^6$     & 1075.5$^\dagger$   & MC$^{2,EC}$ & 1  & 100  & 0 & 10 & 0\\
  & (k5)& 4     & 1.4  & 10    & 10    &  $4\times 10^6$     & 1075.5$^\dagger$     &   1000 & 2& 100 & 0 & 74 & 0\\
 \hline
  \hline
   &     &   &      &    &      &      &  &1st  | 2nd  & 1st  | 2nd  &1st  | 2nd  &1st  | 2nd  & 1st  | 2nd & 1st  | 2nd \\
   \hline
    \hline
 NS-NS  & (l)   
     & 5     & 1.4  & 4     & 1.4   &  3     & 5     &  1000 & 20 |  0  & 0 |  0 &0 |  0 & 0 |  0 & 0 | 7\\
 ($2\times$~SN) & (m)   & 5     & 1.4  & 4     & 1.4   & 3     & 5     & MC$^{2,EC}$ & 20 | 0 & 0 |  0 & 0 |  0 & 0 |  0  & 100 | 0\\
  &(n)  & 5     & 1.4  & 4     & 1.4   & $4\times 10^6$     & 5  & MC$^{2,BW}$ & 20 | 11 & 98 | 43 & 0 | 0 & 0 | 0  & 2 | 57\\
&(o)    & 5     & 1.4  & 4     & 1.4   &  $4\times 10^6$      & 5     & 1000 & 20 | 10 & 100 | 93 &0 | 0 & 3 | 0 & 0 | 7 \\
  &(l5)& 5     & 1.4  & 4     & 1.4   &  3     & 1075.5$^\dagger$     &  1000 & 1 |  0  & 6 |  50 &0 |  0 & 0|  0 & 0 | 0\\
 &(m5) & 5     & 1.4  & 4     & 1.4   & 3     & 1075.5$^\dagger$     & MC$^{2,EC}$ & 1 | 0 & 17 | 50 & 0 |  0 & 21 |  50  & 0 | 0\\
  &(n5) & 5     & 1.4  & 4     & 1.4   & $4\times 10^6$     & 1075.5$^\dagger$  & MC$^{2,BW}$ & 1 | 0.3 & 100 | 100 & 0 | 0 & 9 | 35  & 0 | 0\\
& (o5)& 5     & 1.4  & 4     & 1.4   &  $4\times 10^6$      & 1075.5$^\dagger$     & 1000 & 1 | 0.3 & 100 | 100 & 0 | 0 & 84 | 81  & 0.1 | 0\\
   \hline
  
 BH-BH &(p) & 31    & 30   & 15    & 14    & 3     & 5     &  1000 & 47 | 0 & 0 | 0   & 0 | 0 & 0 | 0  & 100 | 100\\
 ($2\times$~SN) &(q) & 31    & 30   & 15    & 14    & 3     & 5     & MC$^{2,EC}$ & 47 | 0 & 1 | 0 & 0 | 0 & 0 | 0  & 99 | 100\\
  &(r)& 31    & 30   & 15    & 14   & $4\times 10^6$  & 5   & MC$^{2,BW}$ & 47 | 29  & 87 |  83 & 0 | 0 & 0 |  0  &  13 | 17\\
  &(s) & 31    & 30   & 15    & 14   & $4\times 10^6$     & 5     &  1000 & 47 | 32  & 99 | 98 & 0 | 0 & 1 |  0 & 1 | 2\\
    &(p5)& 31    & 30   & 15    & 14    & 3     & 1075.5$^\dagger$     &  1000 & 4 | 0 & 9 | 88   & 0 | 0 & 3 |14  & 0.2 | 0\\
 &(q5) & 31    & 30   & 15    & 14    & 3     & 1075.5$^\dagger$     & MC$^{2,EC}$ & 4 | 0.2 & 14 | 100 & 0 | 0 & 9| 19  &  0.4 | 0\\
  &(r5)& 31    & 30   & 15    & 14   & $4\times 10^6$  & 1075.5$^\dagger$   & MC$^{2,BW}$ & 4 | 2  & 99 |  100 & 0 | 0 & 5 |  4  &  0 | 0\\
  &(s5) & 31    & 30   & 15    & 14   & $4\times 10^6$     & 1075.5$^\dagger$     &  1000 & 4 | 2  & 100 | 100 & 0 | 0 & 54 |  52 & 0 | 0\\
   \hline
\end{tabular}

 \caption{ 
 Table of the numerical experiments run below. We show the masses of the inner binary (pre- and post- SN), the mass of the tertiary, and their SMA. We also present the fraction of systems out of all the runs that remained bound after the SN (column 9), and the fraction of triple systems that remained bound out of all the surviving binaries (column 10). We also show the fraction of systems at which one of the binary members crossed the inner Roche radius ( ($R_{\rm Roche,in}$), see Eq.~(\ref{eq:Roche})) out of all binaries.  The last column shows the fraction of systems of which one of the binary members crossed their tertiary Roche radius ($R_{\rm Roche,out}$, see Eq.~(\ref{eq:a2Cross})) out of all {\it surviving triple} systems. For NS-NS and BH-BH cases we considered two SN explosions.  \\ {\bf MC} represents Monte-Carlo runs, see text and table \label{table:MC} for more details.  The details are specified in the text and for completeness we reiterate our Monte-Carlo initial conditions here. {\bf MC$^1$} refers to the Monte-Carlo choices for $a_1$, which is chosen to be uniform in log space between $5$~R$_\odot$ and $1000$~R$_\odot$. {\bf MC$^{2,EC}$}, refers to the choice of $a_2$, from a uniform in log distribution with a minimum $a_2$ which is consistent with $\epsilon=0.1$ and maximum of $10,000$~AU. The density of binary systems in this case is consistent with $a_2^{-3}$, and thus we label it ''EC'' for extreme cusp. {\bf MC$^{2,BW}$}  refers to the Monte-Carlo choices of $a_2$ to be uniform, which is consistent with density of $a_2^{-2}$ with a minimal value $100$~AU and a maximum value of $0.1$~pc, \citep[which is representative of a distribution around an SMBH, e.g.,][]{Bahcall+76}. Note that the inner and outer SMA also satisfy $\epsilon=0.1$ criteria. In all of our Monte-Carlo runs the inner and outer eccentricities were chosen from uniform distribution, the mutual inclination was chosen from an isotropic distribution. The inner and outer arguments of pericenter and the mean anomaly were chosen from uniform distributions. Interestingly, we note that around 10\% survived inner binaries in BHB systems obtain hyper-velocity. Note that survival rate for binaries and triples refer to the systems that are bound instantaneously post-SNe. The inner binaries that crossed the Roche limit of each other and the binary systems that crossed the Roche limit of the tertiary body are included in the count of survived systems since the systems that are undergoing mass transfer, still stay bound post SNe instantaneously. We provide their percentages in separate columns for clarity.
$^\dagger$ Note that $1075.5$~R$_\odot=5$~au }\label{table E}
\end{table*}

\begin{figure*}[t!]
\begin{center}
\includegraphics[width=\linewidth]{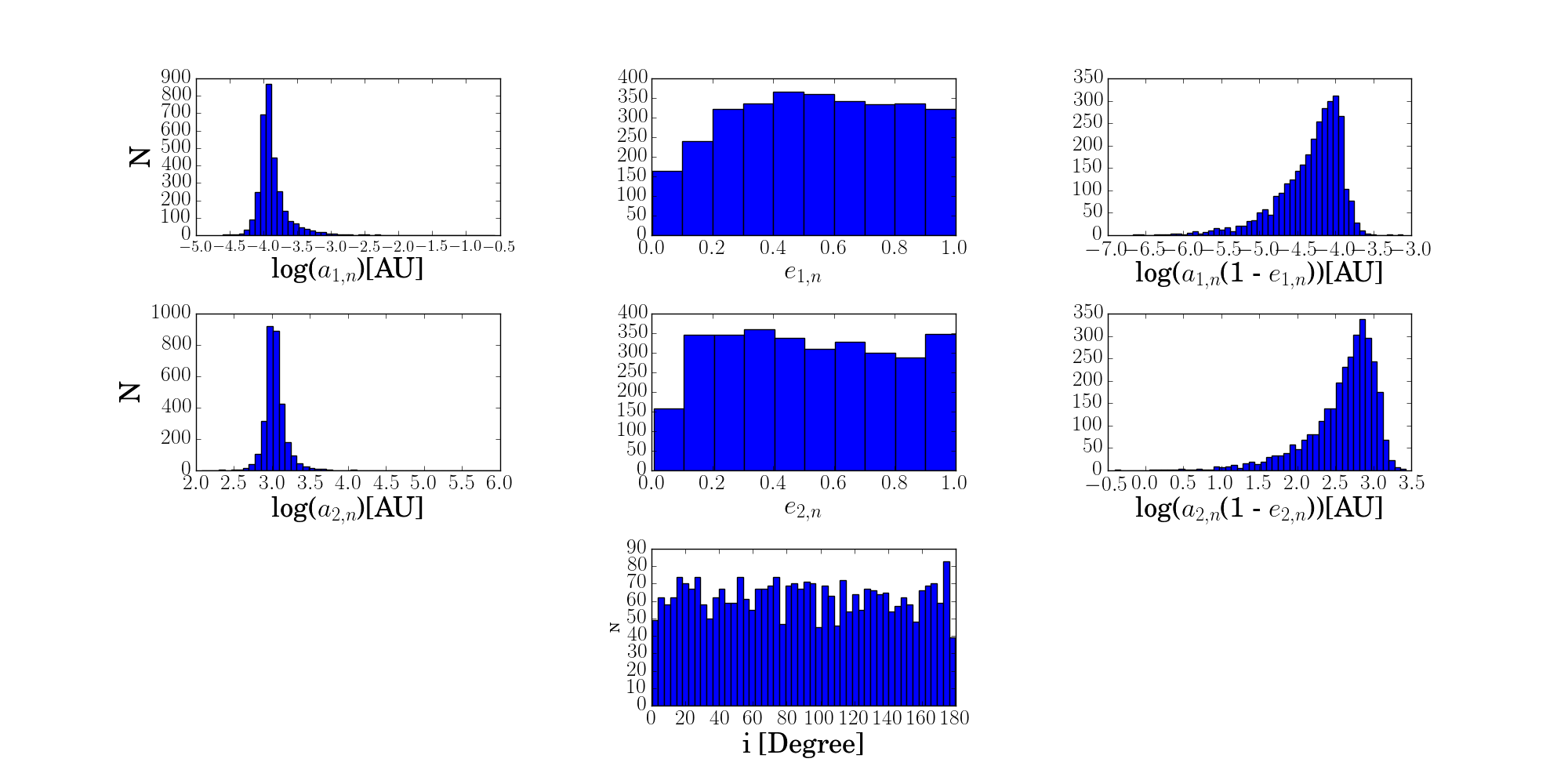}
\caption{\upshape {\bf{The post-SN for BHB orbital parameters distribution after two SN with tertiary SMBH}}. BHB system has initial condition of $a_1 =5$~R$\odot$ and $a_2=1000$~AU. See Section \ref{sec:LIGO-2BH} for description of initial conditions and  Table \ref{table} for statistics. This system represents a typical system investigated by \citep{Hoang+17}. \vspace{0.3cm} }\label{fig:BH_grid}
\end{center}
\end{figure*}

\end{document}

%% file: chapters/introduction.tex
\section{INTRODUCTION}

  The  majority of massive  stars reside in a binary system \citep[$\gsim 70\%$  for OBA spectral type stars; see][]{Raghavan+10}. In addition, observational campaigns have suggested that 
  probably many of these stellar binaries are in fact triples or higher multiple configurations  \citep[e.g.,][]{T97,Pri+06,Tok08,Eggleton+07,Borkovits+16}. From dynamical stability arguments
these must be hierarchical triples, in which the (inner) stellar binary is
orbited by a third star on a much longer orbital period. Therefore, in most cases the dynamical behavior of these systems takes place on timescales much longer than the orbital periods. Recent developments in the study of the dynamics of  hierarchical triples showed that these systems have  rich and exciting behaviors. Specifically, it was shown that the inner orbital eccentricity can reach very high values and the mutual inclination between the two orbits can flip from below $90^\circ$ to above $90^\circ$, namely the Eccentric Kozai-Lidov (EKL) mechanism \citep[see for review][]{Naoz16}. 

Stellar evolution plays an important role in the orbital dynamical evolution of  massive stellar systems  \citep[e.g.,][]{Sana+12}. For example, as the star evolves beyond the main sequence, it losses mass and expand it radius, which can have significant effects on the dynamics of these triple systems \citep[e.g.,][]{Shappee+13,Michaely+14,Stephan+16,Stephan+17,Naoz+16, Toonen2016, Perets2012} . Most notability massive stars ($>8$~M$_\odot$) undergo a Supernova (SN)  explosion of which the star losses a significant fraction of its mass over short mount of time. Including SN-kick to these systems can trigger eccentricity excitations and inclination flips in systems that pre-SN where unfavorable to the EKL mechanism. Of course SN-kicks can also unbind the system, \citep[e.g.,][]{Parker16,Michaely+16}.
  
  Observations of pulsar proper motions in the last decade have shown that neutron-stars receive a large ``natal" kick velocity (with an average birth velocity of $200-500$~km~s$^{-1}$) as a result of SN asymmetry \citep[e.g.,][]{Fryer+99,Lorimer+97,Hansen+97,Cordes+98,Hobbs+04,Hobbs+05,Beniamini+16}. 
   Furthermore,  it was shown that spin--orbit  misalignment in pulsar binary systems requires a natal kick  \citep{Lai+95,Kaspi+96,Kalogera96,Kalogera+98,Kalogera00}.
  The survival of compact object binary systems is extremely  interesting in light of the recent   Laser  Interferometer  Gravitational-wave  Observatory  (LIGO) detection of Black Hole (BH) and NS binary mergers through Gravitational waves emission (GW) \citep[e.g.,][]{LIGO_PRL,LIGO3,LIGO_BH2,Abbott+17NS}.Either of these configurations' progenitors have undergone SN explosion and perhaps even a kick.

  An analytical description of the effect of an SN-kick on a binary system was studied in great details for circular binaries \citep{Hills83, Kalogera96,Tauris1998, Kalogera00,Hurley+02}, mainly for neutron stellar systems. Later, \citet{Belczynski+06} conducted a numerical analysis of black hole (BH) eccentric binaries. Later \citet{Michaely+16} conducted a population synthesis models in binary systems exploring a larger range of parameter space.  Furthermore, recent studies has shown that supernovae in binaries at the galactic center can generate hypervelocity stars \citep{Zubovas2013}. Recently, \citet{Hamers2018} derived a Hamiltonian formalism description to include external perturbations in hierarchical triple systems. These seminal studies showed that SN-kicks play a crucial role in the formation of the spin--orbit  misalignment in Neutron Star (NS) binaries which will affect their gravitational radiation waveforms. Here we expand upon these  works and study the effect of SN-kicks on triple systems, considering a wide range of masses.

  Here we explore SN-kicks in the inner binary of a hierarchical stellar system with an arbitrary inclination and eccentricity\footnote{Note that we neglect the interaction of the supernova ejecta with the companion star since the effect of ejecta-companion interaction is small  \citep[e.g.,][]{Hirai2018}.}. We focus on both triple stellar systems as well as binary stars near an SMBH.
It was recently  suggested that binaries are prevalent in the galactic center. Observationally, there are currently three confirmed binaries in the galactic center \citep[e.g.,][]{Ott+99,Martins+06,Pfuhl+13}. Moreover, it was estimated that the total massive binary fraction in the galactic center is comparable to the galactic binary fraction \citep[e.g.,][]{Ott+99,Rafelski+07}. Furthermore, the recent observations of a gas-like object that plunges toward the Super Massive Black Hole (SMBH) in the center of the galaxy \citep[e.g.,][]{Gillessen+12}, known as G2, provided another piece of evidence for the high likelihood of the existence of young binary systems \citep[e.g.,][]{Witzel+14,Witzel+17,Stephan+16, Bortolas2017}. Theoretically,  \citet{Stephan+16} showed that the binary fraction in the nuclear star cluster might be as high as $70\%$, compared to the initial binary fraction, following a star formation episode that took place in that region a few million years ago \citep[e.g.,][]{Lu+13}. In addition, it was recently suggested that the puzzling observations associated with the stellar disk in the center of our galaxy may provide indirect evidence of a large binary fraction \citep{Naoz+18}.

    Our paper is orgenized as follow: we first describe the set up of our systems (\S \ref{sec:setup}). We then derive the analytical expression for the relevant orbital parameters  (\S \ref{sec:eqs}), and consider few different applications (\S \ref{sec:App}). Specifically, consider applications to potential LIGO sources (\S \ref{sec:LIGO}), specifically, double Neutron star systems (\S \ref{sec:LIGO-2NS}), Neutron star-Black hole binaries (\S \ref{sec:LIGO-NS-BH}) and black hole binaries sources (\S \ref{sec:LIGO-2BH}). We then   we consider SN-kicks in Low mass X-ray binaries (\S \ref{sec:LMXB}), for a Neutron star  (\S \ref{sec:LMXB-NS}) and black hole (\S \ref{sec:LMXB-BH}) compact object. We offer our discussion  in \S \ref{sec:dis}.
  

%% file: chapters/system_setup.tex
\section{System setup}\label{sec:setup}

Throughout this paper we consider the hierarchical triple system which consists of a tight binary ($m_1$ and $m_2$) and a third body ($m_3$) on a much wider orbit. The frame of reference chosen here is the invariable plane defined such that the $z$-axis is parallel to the total angular momentum of the system ${\bf G}_{\rm tot}$ \citep[note that we are using the Delaunay's elements to denote the orbital parameters see][]{3book}. Due to the hierarchical nature of the system the dominant motion of the triple can be reduced into two separate Keplerian orbits: the first describing the relative tight orbit of bodies~$m_1$ and~$m_2$, and the second describes the wide orbit of body~$m_3$ around the center of mass of bodies~$m_1$ and~$m_2$. 
 In this frame we define the orbital parameters, i.e., the  semi-major axes (SMAs) and the eccentricity   of the inner and
outer orbits  as $a_1$, $e_1$ and $a_2$, $e_2$, respectively. 
The inclination of the inner  (outer) orbit  $i_1$ ( $i_2$) is defined as the angle between the inner (outer) orbit's angular momentum ${\bf G}_1$ (${\bf G}_2$)  and the total angular momentum ${\bf G}_\tot$. The mutual inclination is defined as $i_\tot=i_1+i_2$. 

\begin{figure}
\begin{center}
\includegraphics[width=\linewidth]{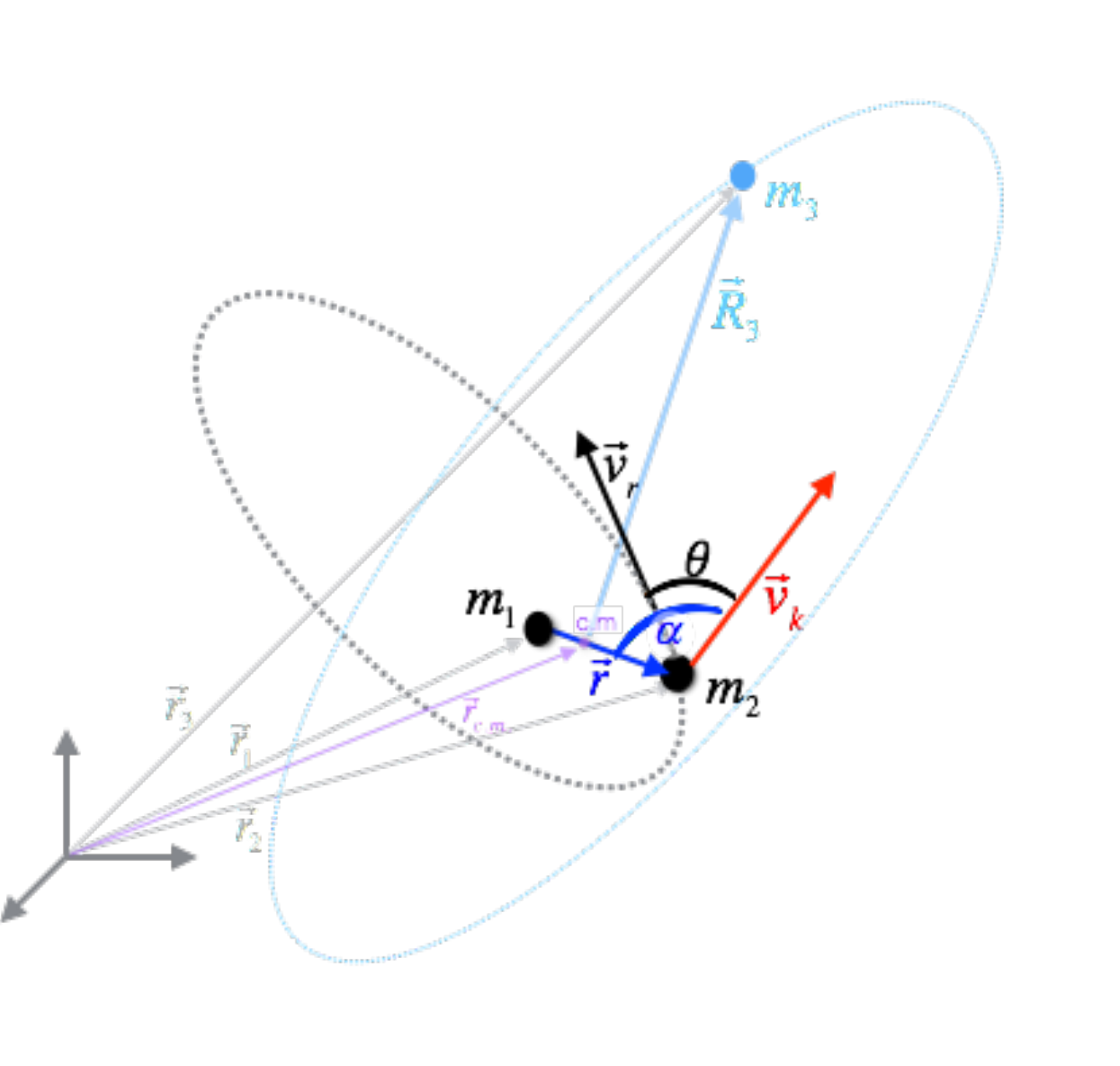}
\caption{\upshape Cartoon of the system (not to scale). Without the loss of generality we chose $m_2$ to undergo SN. See text for details. } \label{fig:Cartoon}
  \end{center}
\end{figure}

Without loss of generality we allow $m_2$ to undergo SN instantaneously, i.e., on a timescale shorter than the orbital period,   associated with a kick velocity ${\bf v}_k=(v_x,v_y,v_z)$. Given a magnitude $v_k$ the direction of the kick velocity vector can be determined by the the angles $\theta$ and $\alpha$ defined such that 
\begin{eqnarray}
{\bf v}_r\cdot {\bf v}_k&=&v_k v_r\cos \theta \\ 
{\bf r}\cdot {\bf v}_k&=&v_k r\cos \alpha
\end{eqnarray}
where ${\bf r}$ and ${\bf v}_r$ are defined in Figure \ref{fig:Cartoon}, with respect  to the plane of the inner orbit\footnote{Note that while the definition of $\theta$ is consistent with that of \citet{Kalogera00}, we choose to define the second angle in a different way. }. 
The magnitude of the position vector is simply:
\begin{equation}\label{eq:rmag}
r=\frac{a_{1}(1-e_{1}^2)}{1+e_{1}\cos f_1} = a_1 (1-e_1\cos E_1)\ ,
\end{equation}
where $f_1$ is the true anomaly of the inner orbit at the time of the explosion. The eccentric anomaly $E_1$ is related to the true anomaly $f_1$ by:
\begin{equation}
\tan\frac{f_1}{2}=\sqrt{\frac{1+e_1}{1-e_1}}\tan\frac{E_1}{2} \ ,
\end{equation}
\citep[e.g.,][]{MD00}.  We will use the eccentric anomaly for our expressions below. The magnitude of the velocity is:
\begin{equation}\label{eq:vmag}
v_r=\sqrt{\mu \left(\frac{2}{r}-\frac{1}{a_1} \right)} = v_c \sqrt{\frac{1+e_1\cos E_1}{1-e_1\cos E_1}} 
\end{equation}
where $v_c=\sqrt{\mu/a_1}$ is the velocity of a circular orbit, and $\mu=G(m_1+m_2)$.  
The scalar product between the orbital velocity and the position vector has a simple relation, using the above definitions,
\begin{eqnarray}
{\bf v}_r\cdot {\bf r} &=& v_r r \cos \eta \\ &=&   v_c r \frac{e_1 \sin f_1}{\sqrt{1-e_1^2}} = v_c a_1{ e_1 \sin E_1} \ . \nonumber
\end{eqnarray}
From this we find that 
\begin{equation}
\cos \eta=\frac{e_1\sin E_1}{\sqrt{1-e_1^2\cos^2E_1}} \ .
\end{equation}
Using $\eta$ the geometry in Figure \ref{fig:Cartoon} yields boundary limits for $\alpha$, i.e., 
\begin{equation}\label{eq:alpha}
\theta-\eta\leq \alpha \leq \theta+\eta \ . 
\end{equation} 

\begin{figure*}[t!]
\begin{center}
\includegraphics[width=\linewidth]{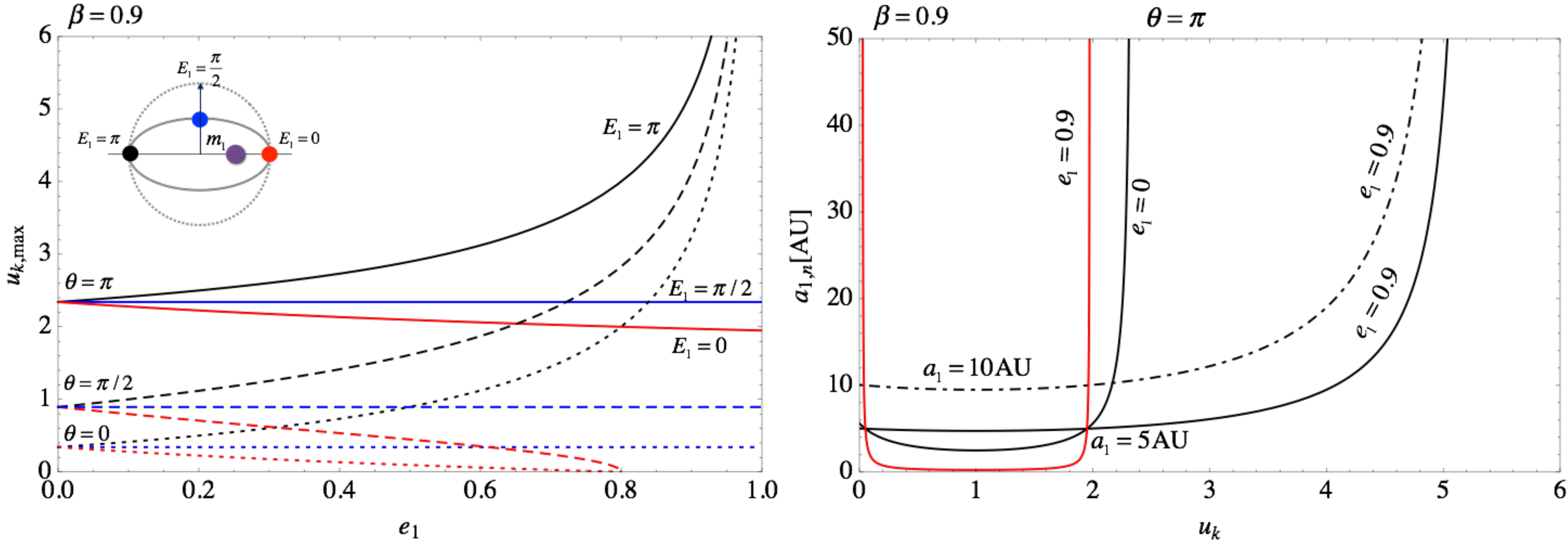}
\caption{\upshape Inner binary orbital parameters,  {\bf left panel:} The maximum dimensionless kick velocity $u_{k,\rm max}={\rm max}(v_k/v_r)$ as a function of the initial inner eccentricity $e_1$, see Equation  (\ref{eq:umax}). We consider three values for the eccentric anomaly $E_1=0,\pi/2$ and $\pi$, red, blue and black lines respectively (see top left cartoon for the orbital configuration). We also consider three possible $\theta$ values (the angle between the kick velocity vector and ${\bf v}_r$), as labled. {\bf Right panel:} The post-SN semi-major axis, $a_{1,n}$ as a function of $u_k$, see Equation (\ref{eq:SMA1}).   We consider initial   semi-major axis $a_1=5$~AU  and show one example for $a_1=10$~AU (dot-dashed line). The color code follows the left panel, i.e., black lines are for $E_1=\pi$ and red line for $E_1=0$. The initial eccentricity is labeled. \vspace{0.3cm}
} \label{fig:umax}
  \end{center}
\end{figure*}

%% file: chapters/post_SN_orbital_param.tex
\section{post-SN Orbital Parameters}\label{sec:eqs}

We consider a system described in Figure \ref{fig:Cartoon} with subscript ``$1$" for the inner orbit and ``$2$" for the outer orbit. 
We denote the post-SN orbital parameters with a subscript ``n".  
The post-SN velocity vector of the new inner orbit is ${\bf v}_{r,n}={\bf v}_{r}+{\bf v}_k$.  The new velocity can be written as: % (see Appendix A)
\begin{equation}\label{eq:vrn}
v_{r,n}^2= \left( {\bf v}_{r}+{\bf v}_k \right)^2=G(m_{1}+m_{2,n})\left(\frac{2}{r}-\frac{1}{a_{1,n}} \right)
\end{equation}
and for instantaneous explosion we have $r=r_n$ \citep[e.g.,][]{Kalogera00} and thus we can solve for the new semi-major axis and find:
\begin{equation}
\label{eq:SMA1}
\frac{a_{1,n}}{a_{1}} = \frac{\beta (1-e_{1}\cos E_1)}{2\beta - (1+e_{1}\cos E_1)(1+u_k^2+2u_k\cos\theta)} \ ,
\end{equation}
where 
\begin{equation}
\beta=\frac{m_1+m_{2,n}}{m_1+m_{2}} \ ,
\end{equation}
the normalized velocity is  $u_k=v_k/v_r$. 
 Note that when $e_{1,0}\to 0$ Equation (\ref{eq:SMA1}), reduced to the relation founds in \citet{Kalogera00}.
 Since the new semi-major axis needs to be positive, it implies that a bound orbit will take place only if the denominator in Equation (\ref{eq:SMA1}) will be positive. In other words, a bound orbit will take place if 
\begin{equation}\label{beta eq}
2\beta > ( 1+ e_1\cos E_1)(1+u_k^2+2u_k\cos\theta) \ .
\end{equation}
Solving for $u_k$ we can find the maximum kick velocity that will allow a bound inner binary. This gives the range of $u_k$ 
\begin{equation}\label{eq:umax}
u_{k, \rm min} \leq u_k  \leq -\cos\theta+\sqrt{\frac{2\beta}{1+e_1\cos E_1}-\sin^2\theta} \ ,
\end{equation}
where 
\begin{equation}
u_{k, \rm min}={\rm max}\left( 0, -\cos\theta\pm\sqrt{\frac{2\beta}{1+e_1\cos E_1}-\sin^2\theta}  \right)  \ .
\end{equation}
As can be seen from this equation, since $1+ e_1\cos E_1$ can range between $0$ and $2$, and $\sin^2\theta\leq 1$ it implies that larger $\beta$ allows for larger range of solutions. The maximum $u_k$ and the post-SN semi-major axis for a nominal choice of initial parameters is depicted in Figure  \ref{fig:umax}. It is interesting to note, as depicted in the Figure, the maximum $u_k$ increases with eccentricity when the SN takes place at apo-center. 
Furthermore, from Equation (\ref{eq:SMA1}), it is clear that if $\beta>1+u_k^2+2u_k\cos\theta$ then the post semi-major axis {\it decreases} with respect to the pre-SN one. In other words, \begin{eqnarray}\label{eq:a1na1}
\frac{a_{1,n}}{a_{1}} 
<1 & \quad\text{if}\quad \beta>1+u_k^2+2u_k\cos\theta \\
\text{and} && \nonumber \\\label{eq:a1na1v2}
\frac{a_{1,n}}{a_{1}}>~1 &\quad\text{if}\quad \beta<1+u_k^2+2u_k\cos\theta
\end{eqnarray} 
We note that the fraction is of course positive at all times. 

An interesting limit can be reached  when $u_k\to 0$, the sudden mass loss shifts the center of mass, and a new semi-major axis can be found by:
\begin{equation}
a_{1,n}(u_k\to 0)=\frac{a_1 \beta  (1 - e_1 \cos E_1)}{2\beta -1  - e_1 \cos E_1} \ .
\end{equation}
The new inner orbit eccentricity can be found from the expression for the angular momentum $h_{1,n}=\sqrt{G(m_{1}+m_{2,n})a_{1,n} (1-e_{1,n}^2)}$, where \begin{equation}\label{eqh1n} h_{1,n}={\bf r}\times {\bf v}_{r,n} \ . \end{equation} Thus, solving for $e_{1,n}$ we find that 
\begin{equation}\label{eq:e1n}
e_{1,n}^2=1-\frac{|{\bf r}\times ( {\bf v}_{r}+{\bf v}_k) |^2 }{ a_{1,n} G(m_1+m_{2,n} )} \ .
\end{equation}
Note that as both the numerator and denominator are proportional to $a_1$, it cancels out and thus $e_{1,n}$ does  not depend on $a_1$.  This yields another conditions for bound orbit, for which $ e_1 \cos E_1>1$.  This has a simple dependency in the orbital parameters, and its a simple equation of $e_1,u_k,\alpha,\theta$ and $E_1$ [see Equation (\ref{eq:J1n})] in Appendix \ref{App:J1}.

Using the eccentricity vector one can simply find the tilt angle between the pre- and post-SN orbital plane, which is associated with the  spin-orbit misalignment  angle, see Appendix \ref{App:tilt} (e.g., Equation (\ref{eq:psi})).

Considering the outer orbit, we define the position vector ${\bf R}_3$ form the center of mass of the inner orbit to the third object (see cartoon in Figure \ref{fig:Cartoon}). The magnitude of this vector is:
\begin{equation}\label{eq:R3mag}
R_3=a_2 (1-e_2\cos E_2) 
\end{equation}

where $E_2$ is the eccentric anomaly 
of the outer orbit at the time the SN in the inner orbit took place. 
 The magnitude of the outer orbit velocity is:
\begin{equation}\label{eq:vmag}
V_{3}=\sqrt{\mu_3 \left(\frac{2}{R_3}-\frac{1}{a_2} \right)} = V_{c3} \sqrt{\frac{1+e_2\cos E_2}{1-e_2\cos E_2}} 
\end{equation}
where $V_{c3}=\sqrt{\mu_3/a_2}$ is the velocity of a circular orbit, and $\mu_3=G(m_1+m_2+m_3)$. Similarly to Equation (\ref{eq:vrn}) we can write the post-SN outer orbit velocity as:
\begin{eqnarray}\label{eq:vR3s}
{V}_{3,n}^2&=& G(m_1+m_{2,n}+m_3)\left(\frac{2}{R_3}-\frac{1}{a_{2,n}}\right) \nonumber \\ 
&=& \left({\bf V}_3-\frac{m_1(m_{2,n}-m_2) {\bf v}_r}{(m_1+m_{2,n})(m_1+m_2)} + \frac{m_{2,n}}{m_1+m_{2,n} }{\bf v}_k \right)^2 \ .
\end{eqnarray}
For the derivation that led to the last transition see Appendix C. This equation can now be used to find $a_{2,n}$. Simplifying it, we can write:
\begin{equation}\label{eq:a2n}
\frac{1}{a_{2,n}}=\frac{2}{a_2 (1-e_2\cos E_2)}-\frac{f_v^2}{G(m_1+m_{2,n}+m_3)} \ ,
\end{equation}
where $f_v^2$ is the right hand side of Equation (\ref{eq:vR3s}) and $f_v^2$ is the right hand side of Eq.~(\ref{eq:vR3s}), i.e,. 
\begin{equation}
 f_v^2= \left({\bf V}_3-\frac{m_1(m_{2,n}-m_2) {\bf v}_r}{(m_1+m_{2,n})(m_1+m_2)} + \frac{m_{2,n}}{m_1+m_{2,n} }{\bf v}_k \right)^2  \ .
\end{equation}
Thus, the constraint that $a_{2,n}\geq0$ can be easily satisfied for a large $m_3$. Interestingly, when $m_3$ is large such that the second term in Equation (\ref{eq:a2n}) goes to zero, the post-SN  kick outer orbit's SMA, $a_{2,n}$, may shrink. 

\begin{table*}
\centering
 \begin{tabular}{l |l || c c c c c c c || c | c | c | c | c} 
 \hline
 Name & Sim &$m_1$ & $m_{1,n}$ & $m_2$    & $m_{2,n}$ & $m_3$     & $a_1$ & $a_2$ &\% Bin & \%  Triples & \% in $R_{\rm Roche,in}$   & \% in $R_{\rm Roche,out}$ & \%escaped \\ 
& & M$_\odot$& M$_\odot$&M$_\odot$ &M$_\odot$  & M$_\odot$ & R$_\odot$    & AU &  out of total & out of Bin &out of Bin & out of 3 &  Bin \\ 
 \hline\hline

  NS-LMXB & (a) & 4     & 1.4  & 1     & 1     & 3     & MC$^1$     & MC$^{2,EC}$ &  4   & 0 & 0 & 0 &  100\\
  & (b) & 4     & 1.4  & 1     & 1     & $4\times 10^6$     & MC$^1$     & MC$^{2,EC}$  &   4 &  94 & 0 & 4  & 6 \\
  & (c) & 4     & 1.4  & 1     & 1     & $4\times 10^6$     & MC$^1$     & MC$^{2,BW}$  & 4  & 99  & 0 &   2 & 1\\
  \hline

 BH-LMXB & (d) & 9     & 7    & 1     & 1     & 3     & MC$^1$    & MC$^{2,EC}$ &   11   & 1 & 24 & 13   & 99\\
 & (e) & 9     & 7    & 1     & 1     &  $4\times 10^6$      & MC$^1$  &   MC$^{2,EC}$ &   11   &  99 & 24 & 7 & 1\\
  & (f) & 9     & 7    & 1     & 1     &  $4\times 10^6$      & MC$^1$  &   MC$^{2,BW}$ &  10   &  92  & 25   & 2 & 8 \\
  \hline

  NS-BH & (g)  & 4     & 1.4  & 10    & 10    & 3     & 5    & 1000 & 33 & 0  & 0 & 0 & 100\\
  & (h) & 4     & 1.4  & 10    & 10    & 3     & 5    & MC$^{2,EC}$ & 33 & 0  & 0 & 0 & 100\\
  & (i) & 4     & 1.4  & 10    & 10    & $4\times 10^6$     & 5    & MC$^{2,BW}$ & 33  & 99  & 0 & 0 & 1\\
 & (j) & 4     & 1.4  & 10    & 10    & $4\times 10^6$     & MC$^1$  & MC$^{2,EC}$ & 12  & 71  & 0 & 2 & 29\\
  & (k)& 4     & 1.4  & 10    & 10    &  $4\times 10^6$     & 5    &   1000 & 33& 100 & 0 & 1 & 0\\
 \hline
  \hline
  &     &   &      &    &      &      &  &1st  | 2nd  & 1st  | 2nd  &1st  | 2nd  &1st  | 2nd  & 1st  | 2nd & 1st  | 2nd \\
  \hline
    \hline
 NS-NS  & (l)   
     & 5     & 1.4  & 4     & 1.4   &  3     & 5     &  1000 & 20 |  0  & 0 |  0 &0 |  0 & 0 |  0 & 0 | 7\\
 ($2\times$~SN) & (m)   & 5     & 1.4  & 4     & 1.4   & 3     & 5     & MC$^{2,EC}$ & 20 | 0 & 0 |  0 & 0 |  0 & 0 |  0  & 100 | 0\\
  &(n)  & 5     & 1.4  & 4     & 1.4   & $4\times 10^6$     & 5  & MC$^{2,BW}$ & 20 | 11 & 98 | 43 & 0 | 0 & 0 | 0  & 2 | 57\\
&(o)    & 5     & 1.4  & 4     & 1.4   &  $4\times 10^6$      & 5     & 1000 & 20 | 10 & 100 | 93 &0 | 0 & 3 | 0 & 0 | 7 \\
  \hline
  
 BH-BH &(p) & 31    & 30   & 15    & 14    & 3     & 5     &  1000 & 47 | 0 & 0 | 0   & 0 | 0 & 0 | 0  & 100 | 100\\
 ($2\times$~SN) &(q) & 31    & 30   & 15    & 14    & 3     & 5     & MC$^{2,EC}$ & 47 | 0 & 1 | 0 & 0 | 0 & 0 | 0  & 99 | 100\\
  &(r)& 31    & 30   & 15    & 14   & $4\times 10^6$  & 5   & MC$^{2,BW}$ & 47 | 29  & 87 |  83 & 0 | 0 & 0 |  0  &  13 | 17\\
  &(s) & 31    & 30   & 15    & 14   & $4\times 10^6$     & 5     &  1000 & 47 | 32  & 99 | 98 & 0 | 0 & 1 |  0 & 1 | 2\\
  \hline
 \end{tabular}
 \caption{ 
 Table of the numerical experiments run below. We show the masses of the inner binary (pre- and post- SN), the mass of the tertiary, and their SMA. We also present the fraction of systems out of all the runs that remained bound after the SN (column 9), and the fraction of triple systems that remained bound out of all the surviving binaries (column 10). We also show the fraction of systems at which one of the binary members crossed the inner Roche radius ( ($R_{\rm Roche,in}$), see Eq.~(\ref{eq:Roche})) out of all binaries.  The last column shows the fraction of systems of which one of the binary members crossed their tertiary Roche radius ($R_{\rm Roche,out}$, see Eq.~(\ref{eq:a2Cross})) out of all {\it surviving triple} systems. For NS-NS and BH-BH cases we considered two SN explosions.  \\ {\bf MC} represents Monte-Carlo runs, see text and table \label{table:MC} for more details.  The details are specified in the text and for completeness we reiterate our Monte-Carlo initial conditions here. {\bf MC$^1$} refers to the Monte-Carlo choices for $a_1$, which is chosen to be uniform in log space between $5$~R$_\odot$ and $1000$~R$_\odot$. {\bf MC$^{2,EC}$}, refers to the choice of $a_2$, from a uniform in log distribution with a minimum $a_2$ which is consistent with $\epsilon=0.1$ and maximum of $10,000$~AU. The density of binary systems in this case is consistent with $a_2^{-3}$, and thus we label it ''EC'' for extreme cusp. {\bf MC$^{2,BW}$}  refers to the Monte-Carlo choices of $a_2$ to be uniform, which is consistent with density of $a_2^{-2}$ with a minimal value $100$~AU and a maximum value of $0.1$~pc, \citep[which is representative of a distribution around an SMBH, e.g.,][]{Bahcall+76}. Note that the inner and outer SMA also satisfy $\epsilon=0.1$ criteria. In all of our Monte-Carlo runs the inner and outer eccentricities were chosen from uniform distribution, the mutual inclination was chosen from an isotropic distribution. The inner and outer arguments of pericenter and the mean anomaly were chosen from uniform distributions. Note that survival rate for binaries and triples refer to the systems that are bound instantaneously post-SNe. The inner binaries that crossed the Roche limit of each other and the binary systems that crossed the Roche limit of the tertiary body are included in the count of survived systems since the systems that are undergoing mass transfer, still stay bound post SNe instantaneously. We provide their percentages in separate columns for clarity.
}\label{table}
\end{table*}

%================================================================================
%================================================================================
Similarly to Equation (\ref{eq:e1n}), we can find the post-SN outer orbit eccentricity 
\begin{eqnarray}\label{eq:e2n}
e_{2,n}^2&=&1-\frac{1 }{ a_{2,n} G(m_1+m_{2,n}+m_3 )} \times \\ 
&&  \bigg|{\bf R}_3\times\left({\bf V}_3-\frac{m_1(m_{2,n}-m_2) {\bf v}_r}{(m_1+m_{2,n})(m_1+m_2)} + \frac{m_{2,n}}{m_1+m_{2,n} }{\bf v}_k \right)   \bigg|^2\ . \nonumber
\end{eqnarray}
The total angular momentum is simply \begin{equation}{\bf G}_{\tot,n}={\bf G}_{1,n}+{\bf G}_{2,n} \ ,\end{equation}  where \begin{equation}{\bf G}_{1,n}=\frac{m_1 m_{2,n}}{m_1+m_{2,n}}{\bf h}_{1,n} \ , \end{equation}
and ${\bf h}_{1,n}$ is defined in Equation (\ref{eqh1n}), and similarly we can define ${\bf h}_{2,n} = {\bf R}_3 \times {\bf V}_{3,n}$ and ${\bf G}_{2,n}$. From these we can find the new mutual inclination (after transferring to the invariable plane):
\begin{equation}
\cos i_{n} = \frac{ G_{\tot,n}^2 - G_{1,n}^2 - G_{2,n}^2 }{2 G_{1,n} G_{2,n} } \ .
\end{equation}
From the angular momentum vectors we can also deduce the line of nodes, and from there, using the eccentricity vectors of the inner and outer orbits one can infer the argument of periapsis of the inner and outer orbit SMA. 

Note that similar equations have been derived in previous literature \citep[e.g.][]{Pijloo2012, Toonen2016, Hamers2018}. We present derived equations here so that our notations can be self-contained.
\begin{figure}[t!]
\begin{center}
\includegraphics[width=\linewidth]{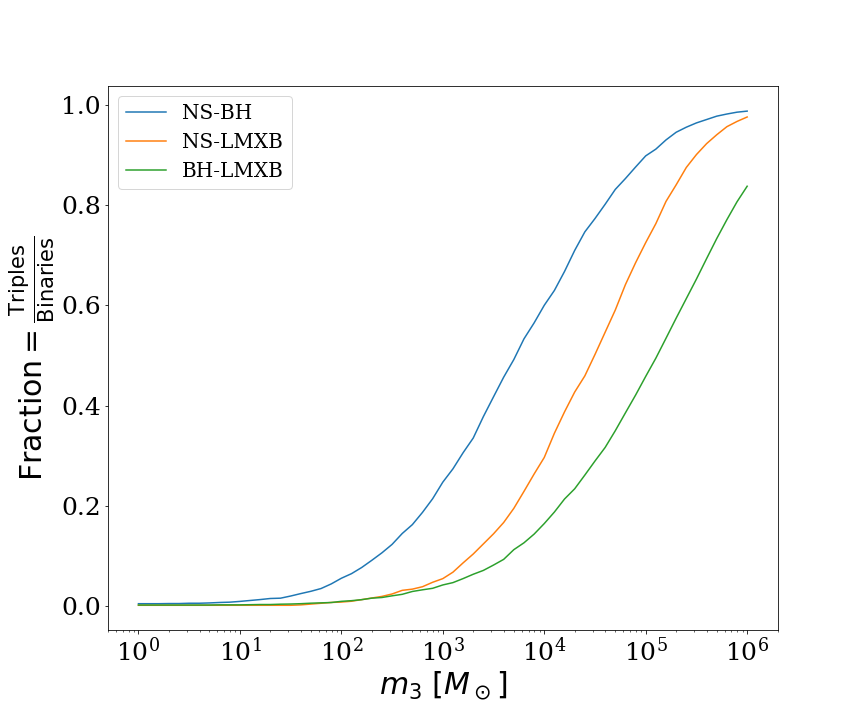}
\caption{\upshape We show the fractions of triples over binaries that survived a NS-BH(NS-LMXB, BH-LMXB) one natal kick as a function of the tertiary mass. This is for systems of NS-BH with $a_1$ = ~$5 R_\odot$ and $a_2$ from MC$^{\text{2,BW}}$. For NS-LMXB and BH-LMXB, $a_1$ is chosen from MC$^{1}$ and $a_2$ from MC$^{\text{2,EC}}$.
\vspace{-0.3cm} }
\end{center}\label{fig:mass}
\end{figure}

%% file: chapters/applications.tex
\section{Applications}\label{sec:App}

We present several representative numerical experiments aimed to explore a variety of astrophysical applications. These are meant to give a proof-of-concept for the types of possible outcomes. We note that the final result depends on the choice of initial conditions, and that a full population synthesis or detailed Monte-Carlo are beyond this initial scope of the paper. The numerical parameters are summarized in Table \ref{table}. In all of our numerical analyses below we work in the invariable plane.
In all of our {\bf Pre-SN} systems  we require a hierarchical  system that satisfied the stability condition: 
\begin{equation}\label{eq:epsi}
    \epsilon = \frac{a_1}{a_2}\frac{e_2}{1-e_2^2} \leq 0.1 \ ,
\end{equation}
\citep[e.g.,][]{LN}.

All of our pre-SN systems are beyond tidal disruption limit, i.e., $a_2 (1-e_2) > a_1 (1+e_1) (m_3/(m_1+m_2))^{(1/3)}$.   Furthermore, note that \citet{Naoz+13}, showed that the $\epsilon$ criterion has a similar functional form as the \citet{Mardling+01}. We also checked that the \citet{Mardling+01} criterion is satisfied for all of our similar mass systems (for which this criterion was devised).  Stability of mass hierarchy was studied in the literature in great details, \citep[e.g.][]{Antonini+14, Antognini+13,Ivanov+05,Katz+12,Bode+14,Hamers+13,Petrovich152p} , and indeed we have verified that not only the binaries around the SMBH are above the tidal disruption zone, and $\epsilon >0.1$ but also that they obey the stability criterion \citep[e.g.][]{Petrovich152p} . We note these stability criteria deem a system unstable if at any point in time it will encounter instability and does not take timescale into consideration \citep[e.g.][]{Myllari18}. Thus, we stress that using these criteria underestimate the number of allowable systems within their lifetime.

We note that we do not provide a population synthesis here, however, we estimate the quadruple-order timescales of our systems to estimate how likely it is that they have underwent a EKL evolution before the SN-kick took place. The systems will not be affected by secular effects before they undergo supernova. We stress that this is a heuristic calculation because a self-consistent one would need to include both the post-main sequence stellar evolution effect on EKL which is beyond the scope of this paper \citep[see for the dramatic implications of the interplay between EKL and post-main sequence evolution, e.g.,][]{Stephan+16, Stephan+17, Stephan+18, Naoz16, Toonen2016}.

We focus our discussion on the survival of the inner binary and the overall triple configurations, as well as the possible outcomes. A kick can unbind the binary or the triple, or, to be more precise, if the post-SN orbital velocity is larger than the escape velocity of the system, the system becomes unbound. Note that the fraction of binaries surviving the SN-kick is of course independent on the  choice of tertiary companion. Although in the Solar neighbourhood the most massive star is the tertiary in about 18 percent of triples \citep{Tokovinin2010}, we only focus on the scenario in which the inner binary undergo SN first and when tertiary goes SN, it will not affect the parameters of inner orbit. Tertiary companion's SN affect on inner orbit is beyond the scope of this paper.

\begin{table}
\centering
 \begin{tabular}{| l |c | c |}
  \hline
  Name & $a_1$ & $a_2$ \\
   \hline
    \hline
  {\bf MC$^1$} &  Uniform &  -  \\
              &  $5-1000$~R$_\odot$ & \\
               \hline
                {\bf MC$^{2,BW}$}& -  &    $n(a_2)\sim a_2^{-2}$ \\
              Bahcall- Wolf         &    & $100$~AU - $0.1$~pc\\
                   \hline
  {\bf MC$^{2,EC}$} & - &        $n(a_2)\sim a_2^{-3}$  \\
            Extreme-Cusp         &   &  $a_{2,{\rm min}}\in \epsilon=0.1$ \\
                    &   &   $a_{2,{\rm max}}=10^4$~AU \\
                     \hline
   
 \end{tabular}
 \caption{A summary of the Monte-Carlo parameters. For the inner orbit's SMA, $a_1$, Monte-Carlo simulations were chosen to be uniform, with the limits specified in the table. For the outer orbit we have followed \citet{Hoang+17} Monte-Carlo choice of stellar distribution around an SMBH.  Specifically, we have  chosen a Bahcall-Wolf \citep[(BW) e.g.,][]{Bahcall+77} distribution as well as an extreme cusp distribution. Note for $N$ number of object we have  $\text{dN} = 4\pi~a_2^2~n(a_2)~\text{d}a_2 = 4\pi~a_2^3~n(a_2)~\text{d}(\ln a_2)$, and thus we choose to have the outer binary SMA follow a uniform distribution in $a_2$ for the BW distribution (MC$^{2,BW}$) and $\ln a_2$  Extreme cusp (MC$^{2,EC}$) distribution. We choose both the inner and outer orbit eccentricities, $e_1$ and $e_2$, respectively,  from a uniform distribution between $0-1$, the argument of perihapsis of the inner and outer orbits, $\omega_1$ and $\omega_1$ respectively,  from a uniform distribution between $0^\circ-180^circ$, the mutual inclination was chosen from an isotropic distribution (uniform in $\cos i$). See text for more details.  }\label{table:MC}
\end{table}

In all of our runs, we assume a normal distribution of kick velocities with an average of $400$~km~s$^{-1}$ and a standard deviation of $265$~km~s$^{-1}$. \citep[e.g.,][]{Hansen+97,Hobbs+04,Arzoumanian+02}. The tilt angle $\theta$ is chosen from a uniform distribution between 0 and $2\pi$ and $\alpha$ is then chosen from a  uniform distribution for which the minimum and maximum values are set by Equation (\ref{eq:alpha}). Furthermore,  in all of our runs, the inclination angle was chosen to be isotropic (uniform in $\cos i$)  and the arguments of periapsis of the inner and outer orbits were chosen from a uniform distribution between $0$ and $2\pi$. The inner and outer eccentricities, $e_1$ and $e_2$, were chosen from a uniform distribution between $0-1$. The mean anomaly was chosen from a uniform distribution from which we solved for the true and eccentric anomalies \citep[e.g.][]{Savransky+11}. See Table \ref{table:MC} for details of the Monte-Carlo simulations and how they depend on the semi-major axis. 
We run a total of $10,000$  systems for each tertiary mass.

\begin{figure}[t!]
% \begin{center}
\includegraphics[width=1\linewidth]{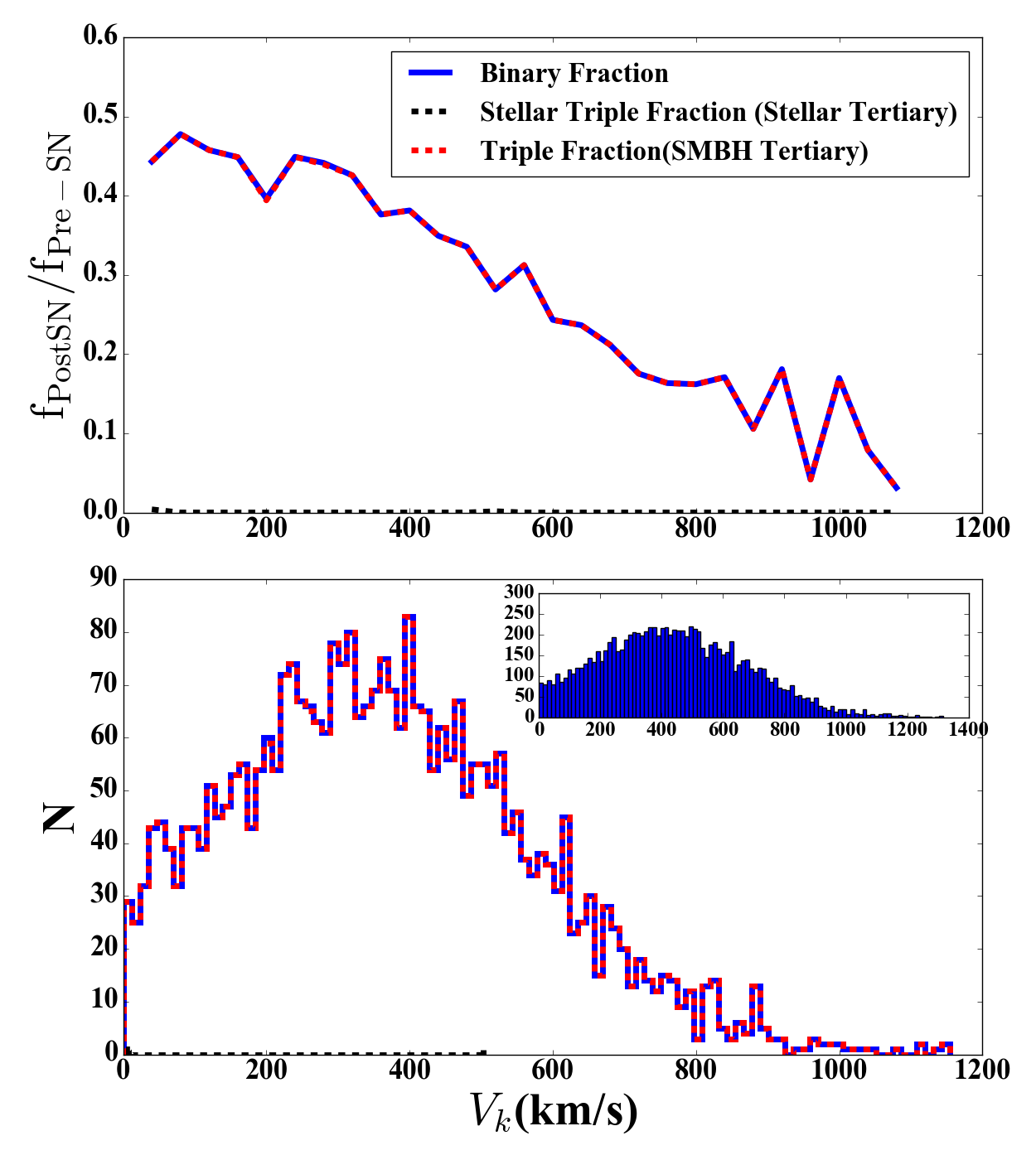}
\caption{\upshape Kick velocity distribution of survived system in {\bf NS-BH binaries}, one natal kick for $a_1=5 R_\odot$ and $a_2=1000$AU, for stellar companion run (g) left. and SMBH companion run (k), right. In the {\bf{ top panel}} we consider the  systems that remained bound post-SN with a distant stellar mass companion in black dotted line and with a distant SMBH companion with red dotted line. Note that the triple fraction is very low (nearly zero) with stellar companion.{\bf Bottom panels:} show a histogram of the kick velocity of the survived binaries (solid lines) and triples (dashed lines). Note that the  survived inner binaries are independent of the mass of the outer body. The inset in the bottom panel shows the initial distribution of the kick velocity $v_k$. Also note that in the case with SMBH as tertiary, the two lines overlap. 
}\label{fig:survived_nsbh}
% \end{center}
\end{figure}

\begin{figure}
%\begin{center}%\hspace{-2cm}
\includegraphics[width=1\linewidth]{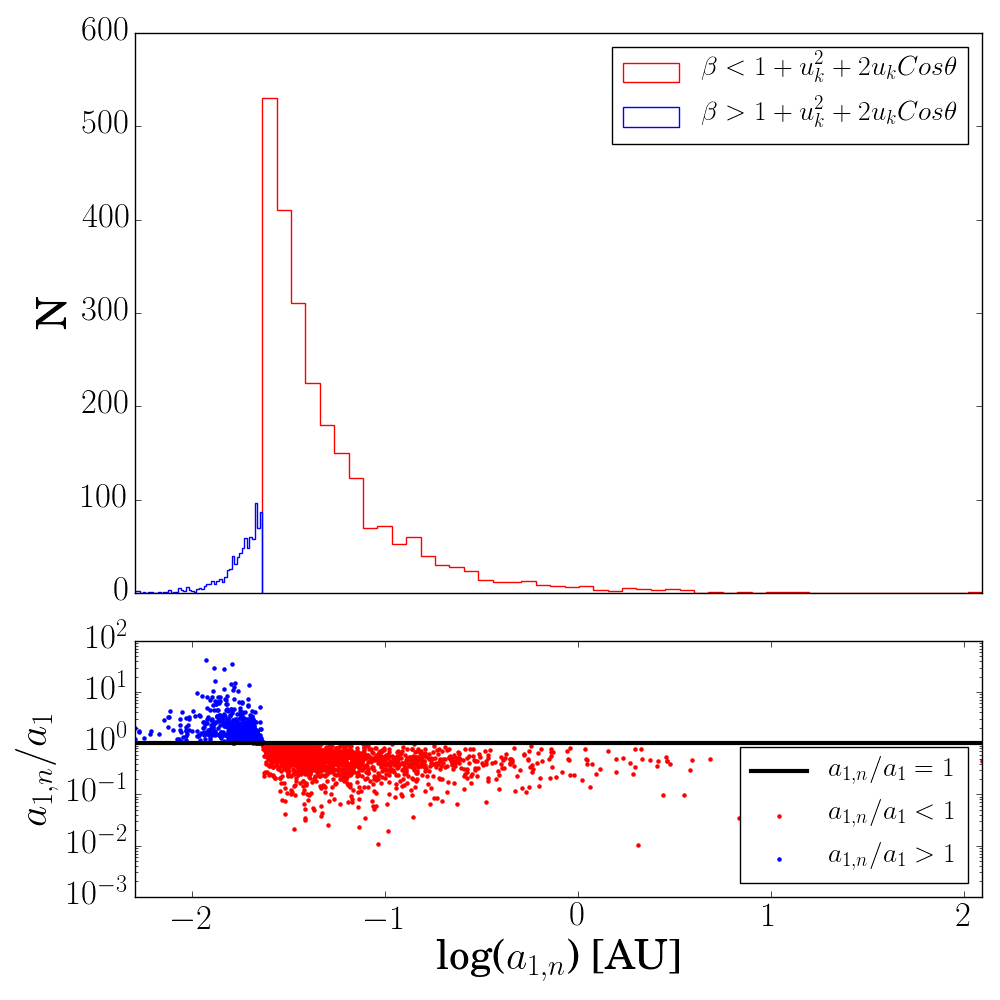} \caption{\upshape  {\bf NS-BH Inner binary Semi-major Axis} for system with $a_1=5 R_\odot$ and $a_2$ from MC$^{\text{2,BW}}$, and an SMBH companion, run (i). We show a histogram ({\bf top panel}) of the inner binary semi-major axis after a SNe in the NS-BH system. We consider systems that their orbit expanded after the two SNs (red line), which corresponds to the last step $\beta<1+u_k^2+2u_k\cos\theta$, as well as systems that shrunk their orbits due to the SNs natal kicks (blue line, $\beta>1+u_k^2+2u_k\cos\theta$). In the {\bf bottom panel} we show the semi-major axis ratio ($a_{1,n}/a_1$) as a function of the post-SN semi-major axis. }\label{fig:NSBH_a1}
\end{figure}

\begin{figure}[t!]
\begin{center}
\includegraphics[width=1\linewidth]{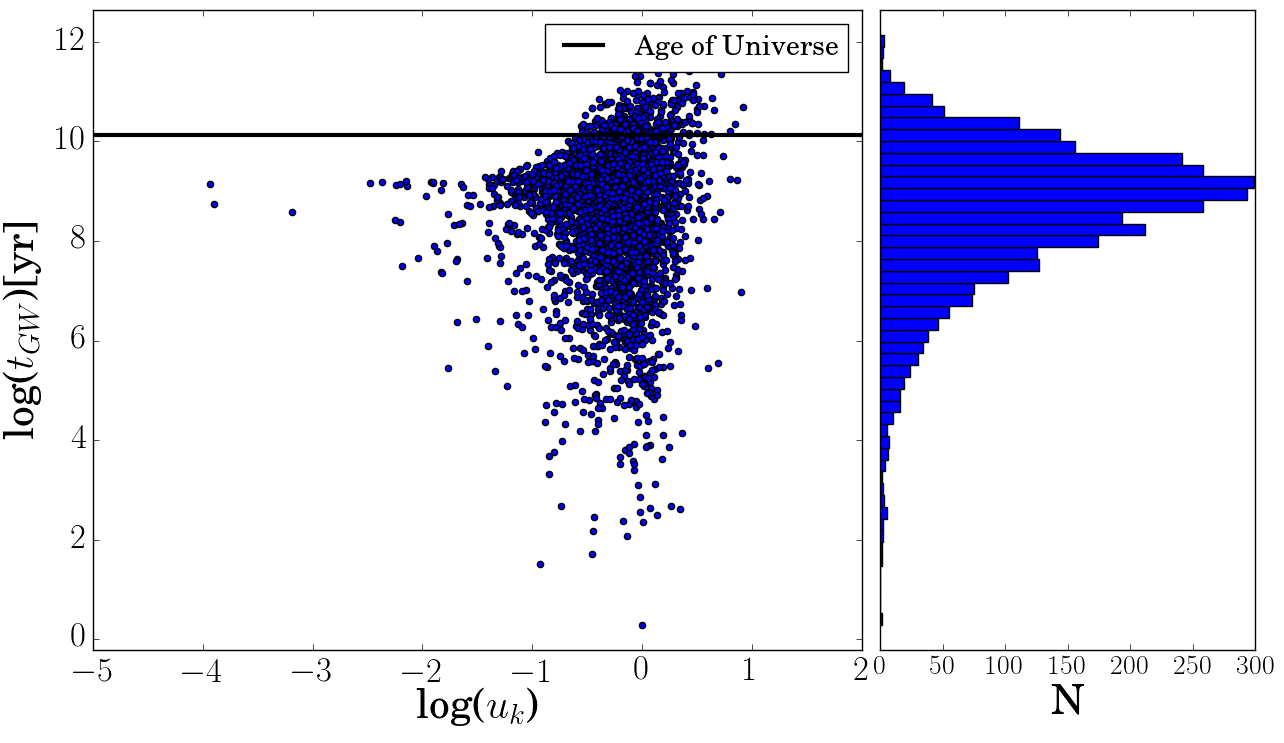}
\caption{\upshape NS-BH GW timescale for system with $a_1=5 R_\odot$ and $a_2$ from MC$^{\text{2,BW}}$, and an SMBH companion, run (i). {\bf{Left Panel}}: We show the GW timescale as a function of the dimensionless kick velocity,  $u_k$. In the  {\bf{right Panel}} we show the histogram of the GW timescales. \vspace{0.3cm} }\label{fig:nsbh_gwt}
\end{center}
\end{figure}

\begin{figure*}[t!]
\begin{center}
\includegraphics[width=\linewidth]{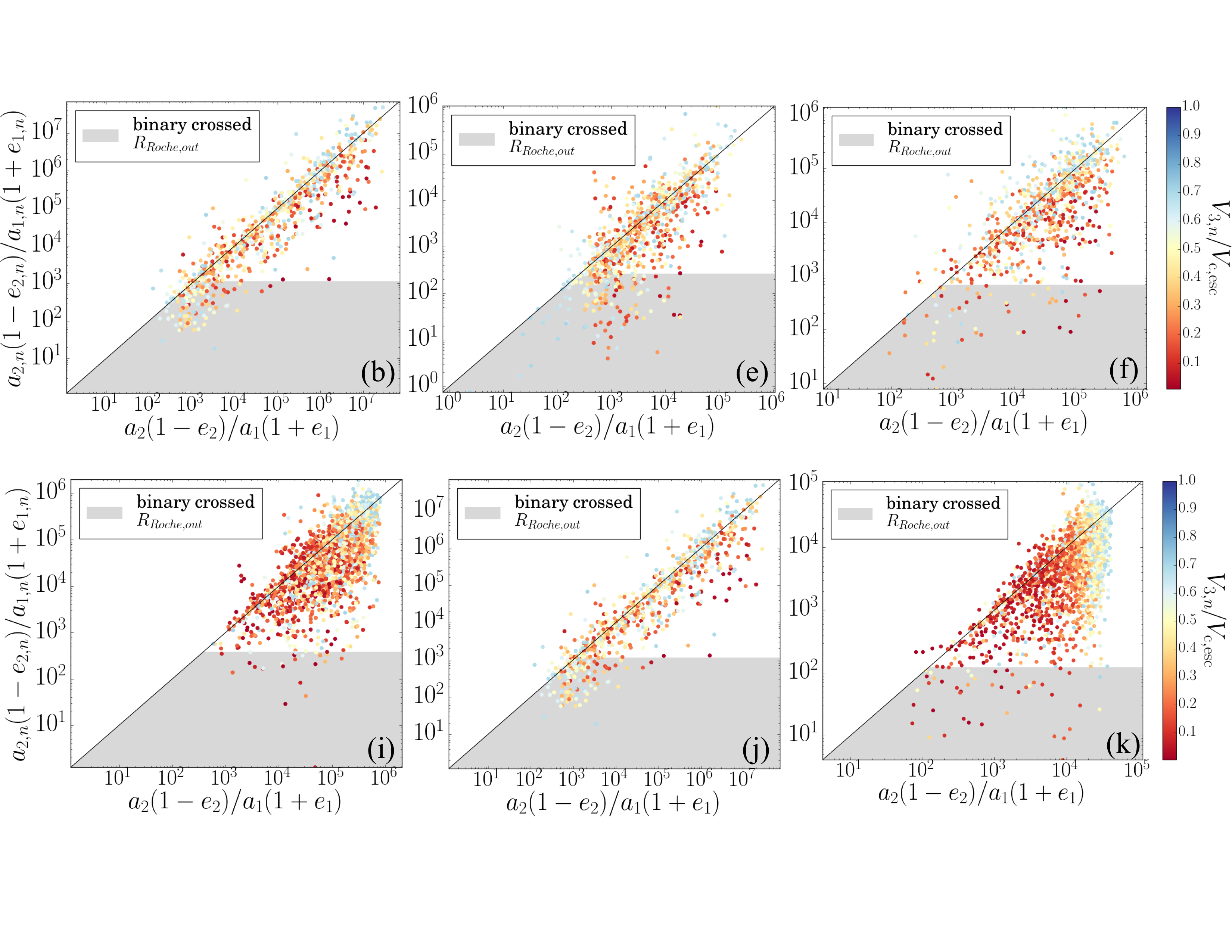}\vspace{-1.5cm}
\caption{\upshape SMBH Roche-limit crossing parameter space for systems after the first (or only) SN. Considering Equation (\ref{eq:a2Cross}) we show the pre- and post- SN parameters. Marked in grey shade are systems that crossed the SMBH Roche-limit and thus, unbind the binary are potential LISA events. We denote in the bottom of each panel the corresponding Monte-Carlo  simulation used. Not that this is  a plot only for the  Monte-Carlo for both $a_1$ and $a_2$. %\vspace{0.3cm}
}\label{fig:a2Cross6}
\end{center}
\end{figure*}

\begin{figure*}[!h]
\includegraphics[width=1\linewidth]{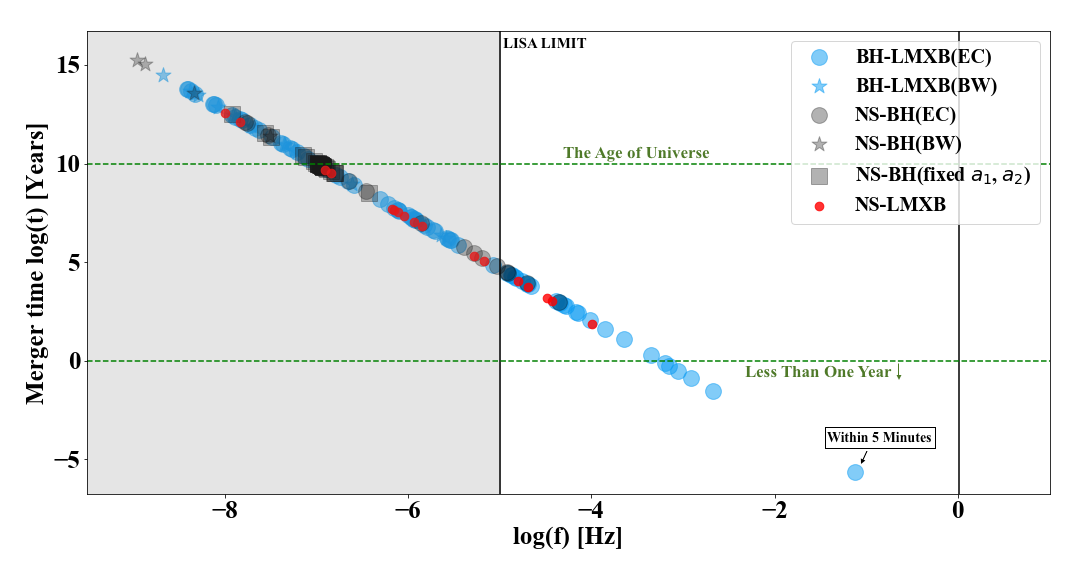} \caption{\upshape  {\bf EMRI GW merger time after SN versus GW emission characteristic frequency}. We consider all systems in our proof-of-concept Monte-Carlo, see below for details, that crossed the SMBH Roche limit (see Eq.~(\ref{eq:a2Cross})). The GW characteristic frequency is computed according to Eq.~(\ref{eq:f}). The black varietal lines are LISA frequency detection limits.  See Table \ref{table} for the fraction of these systems from each Monte-Carlo. {\bf We note that the systems depicted here are only for the $5$~R$_\odot$ runs. The $5$~au runs all resulted in longer than Gyr timescale.  }  %\ref{eq:}
}\label{fig:LISA_prediction} \vspace{-1.3cm}
\end{figure*}

\begin{figure*}
\begin{center}
\includegraphics[width=\linewidth]{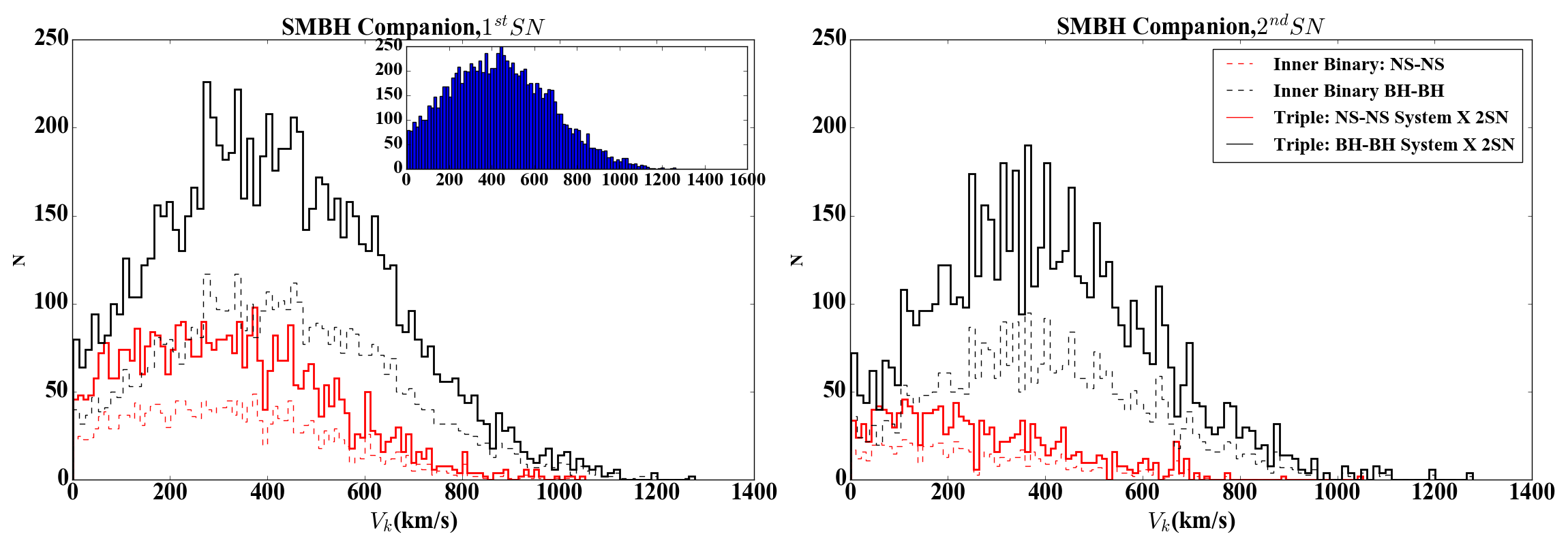}
\caption{\upshape {\bf Two natal kicks case}. Kick Velocity distribution of Survived System in GW Sources after the {\it first} SN (left) and the {\it seconds} SN (right) took place, for BH-BH (black lines) and NS-NS (red lines) systems. We consider systems around and SMBH, with $a_1=5 R_\odot$ and $a_2$ from MC$^{\text{2,BW}}$, corresponding to runs (n) and (r). We show the surviving binaries (dashed lines) and the surviving triples (solid lines) in each case. The inset shows the initial kick velocity  distribution. 
\vspace{0.3cm} }\label{fig:survived_2sn}
\end{center}
\end{figure*}

%% file: chapters/GW_source/ns_bh.tex
\subsection{GW Sources }\label{sec:LIGO}
\subsubsection{Neutron Star - Black hole Binary   - one natal kick}\label{sec:LIGO-NS-BH}

The formation scenario of NS-BH binary systems typically involves that after the first SN  explosion, the compact remnant enters a common-envelope phase with its companion. This may lead to tightening of the orbit, and if the system remains bound after the companion star collapses, a NS-BH binary may form \citep[e.g.,][]{Fryer+99,Dominik+12,Postnov+14}. These compact object binaries have been suggested to exist in the Galactic disk \citep[e.g.,][]{Pfahl+05,Kiel+09}, Galactic center \citep[e.g.,][]{Faucher+11} and globular clusters \citep[e.g.,][]{sig03}.
Recently, \citet{LIGO_NS}  constrained the NS-BH binaries merger rate to be less than  $3,600$~Gpc$^{-3}$~yr$^{-1}$ based on the non-detection so far. 

As a proof-of-concept we run a Monte-Carlo numerical experiment, exploring the possible outcomes of triple systems. For these systems we assume that {\it only} the NS had a natal kick. We adopt \citet{Kalogera00} orbital parameters for the binary, with $m_1=10$~M$_\odot$ and $m_2=4$~M$_\odot$, and a post-SN mass of $1.4$~M$_\odot$. We initially set $a_1=5$~R$_\odot$. However, unlike \citet{Kalogera00}, who adopted a circular initial (inner) orbit, we adopt a uniform eccentricity distribution between $0$ and $1$. We explore two systems with two different tertiaries, one with a stellar companion $m_3=3$~M$_\odot$ and the other with $m_3=4\times 10^6$~M$_\odot$. 
%The latter is coincident with \citep{Faucher+11}.
In both examples we set $a_2=1000$~AU and adopt a uniform distribution for $e_2$, while keeping the stability requirement specified in Equation (\ref{eq:epsi}).

In addition to these two runs, we also run a Monte-Carlo run for the stellar companion by drawing $a_2$ from a distribution with a maximum of $1000$~AU and a minimum $a_2$ which satisfy $\epsilon=0.1$.  For the stellar-companion case we also run a Monte-Carlo run for which $a_2$ is chosen from a log normal distribution with a minimum that satisfied $\epsilon=0.1$ and a maximum of $10,000$~AU. See Table \ref{table}, for a summary of the runs and outcomes and see Table \ref{table:MC} for the Monte-Carlo parameters. 

Furthermore, we also run the same set of Monte-Carlo runs while setting $a_1=5$~AU to allow for a wider initial configuration (see Appendix \ref{App:5au} and   Table \ref{table E} for the parameters). For these systems we found that in this cause a higher fraction of systems crossed the SMBH.

While the fraction of binaries surviving the SN-kick is of course independent of the  choice of tertiary, the  survival  of triples increase with the mass of the tertiary, as depicted in Figure \ref{fig:mass}. As shown in this Figure, the fraction of triple systems that remain bound after the SN occurred approaches the binary fraction for tertiaries with masses {$\gsim 10^6$~M$_\odot$}. Note that this is a generic conclusion for all of our cases. Thus, in our SMBH companion case the fraction of systems that remain bound triples approaches the binary fraction (as depicted in the bottom panels of  Figure \ref{fig:survived_nsbh}). This implies that mergers of compact binaries due to eccentricity induced dynamical evolution such as EKL is more likely to take place in the presence of high-mass third companion \citep[e.g.,][]{Hoang+17}.

In Figure  \ref{fig:survived_nsbh}, we show the dependency of the surviving systems on the kick velocity. In particular, we depict in the bottom panels, the distribution of the surviving binaries (triples), solid (dashed) lines and in the top panels, we show the fraction of surviving systems out of the initial systems. We consider our two tertiary systems, stellar companion (left column) and SMBH companion (right column).
As expected the fraction of systems that stay bound post-SN  decreases as a function of the kick velocity induced, as seen in the top panel of Figure \ref{fig:survived_nsbh}. Specifically, in this Figure we show the fraction of bound post-SN systems  ($f_{\rm PostSN}$) over the fraction of initial pre-SN systems ($f_{\rm Pre-SN}$) at the same kick velocity bin. About $50\%$ of the binaries remained bound for kick velocities $\lsim 300$~km~s$^{-1}$. Furthermore, about $0.3\%$ ($50\%$) of the triples remain bound for stellar mass (SMBH) tertiary, and for the same kick velocity range. In the bottom panels we show a histogram of the bound systems as a function of $v_k$.   

Our analytical calculation showed that systems for which $\beta>1+u_k^2+2u_k\cos\theta$ will shrink their SMA after the natal kick. As depicted in that Figure \ref{fig:NSBH_a1}, given the mass ratio, $\beta$, the tilt angle $\theta$ and the dimensionless kick velocity $u_k$, we can predict if the orbit will shrink or expand.

Shrinking SMA following natal kick can have interesting consequences. For example, NS-BH binaries will merge by emitting GW emission \citep[e.g.,][]{PetersMathews1963, Peters1964, PressThorne1972} and here we use a scaling relation give by \citet{Blaes2002}: 
\begin{eqnarray}
t_{gw}\approx && 2.9 \times 10^{12} \text{yr} \left( \frac{m_1}{10^6 M_{\odot}} \right)^{-1} \left( \frac{m_{2,n}}{10^6 M_{\odot}} \right)^{-1}\\
&&\times \left( \frac{m_1 + m_{2,n} }{2 \times 10^6_{\odot}}\right)^{-1}\left( \frac{a_{1}}{10^{-2} \text{pc}} \right)^{-1} \times f(e_1) \left(1 - e_1^2\right)^{7/2}\ . \nonumber
\end{eqnarray}
We estimate the GW timescale at which these binaries will merge for one of our proof-of-concept runs (see Figure \ref{fig:nsbh_gwt}), and find an average merging time of $\sim 10^{8.5}$ yrs (for $a_1=5$~R$_\odot$ and $a_2=1000$~AU system with an SMBH companion, i.e., run (i)). We find somewhat shorter merger timescale ($\sim 10^6$~yr) of a Monte-Carlo that considers  an initial SMA binary of  $a_1=5$~R$_\odot$ and $a_2=1000$~AU system with stellar companion, i.e., run (g). As depicted in Figure \ref{fig:nsbh_gwt}, the merger times are not overly sensitive to the dimensionless kick velocity $u_k$.  This is consistent with our results and emphasize the sensitivity to initial conditions. Furthermore, we expect that near SMBH the merger time will shorten due to the Eccentric Kozai-Lidov mechanism \citep[e.g.,][]{Naoz16,Hoang+17}.  

We  found that $\sim 1-2\%$  of all surviving triples in the Monte-Carlo runs with SMBH tertiary crossed its Roche limit. 
In other words the resulting post-SN orbital parameters were:
\begin{equation}\label{eq:a2Cross}
a_{2,n}(1-e_{2,n}) < a_{1,n}(1+e_{1,n})\left( \frac{m_3+m_{2,n}+m_1}{m_1+m_{2,n} } \right)^{1/3}
\end{equation}
\citep[e.g.,][]{Naoz+14Silk}. 
 This process will break up the binary similarly  to the Hills process \citep[e.g.,][]{Hills88,Yu+03}. In this case we have that one of the compact object will be on a close eccentric orbit around the SMBH, spiraling in via GW emission (i.e., Extreme Mass Ratio in spiral, EMRI). In particular, we note that for the wider initial binaries the SN-kick results in Roche-limit crossing between one member of the inner tight binary and the tertiary. For the most dramatic cases, the percentage of {\it mergers with the SMBH post-kick}  is $1$\% (for $a_1=5$~R$_\odot$) and $\sim 74$\% (for $a_1=5$~au), see \ref{table E} case (k) and (k5). These events can result in EMRI via GW emission which may be a LISA source.  In Figure, \ref{fig:a2Cross6} we show the ratio of the left to the right hand side of Eq.~(\ref{eq:a2Cross}), up to the mass term, pre- and post SN. We show the results for three of the proof-of-concept Monte-Carlo runs we conducted around SMBH, (i), (j) and (k), see Table \ref{table}. As depicted in all three cases, a non-negligible fraction of the systems crossed the SMBH Roche-limit (shaded grey in the Figure). Thus, resulting in having one member of the binary gaining high velocity and the other on a tight eccentric configuration which can spiral in via GW emission.  

 \begin{table}[h!]
	\centering
 	\begin{tabular}{l |l || c} 
 	\hline
 Name & Sim & $t_{\mathrm{gw}}$[years] crossed SMBH $R_{\text{Roche}}$\\ 
 \hline\hline
  NS-LMXB & (b) & a few ($\sim$1-10) million years\\
 BH-LMXB & (e) &  a few million years \\
  NS-BH & (j)  &  a few million years\\
  \hline
   &     &  1st SN  $t_{\mathrm{gw}}$ | 2nd SN $t_{\mathrm{gw}}$\\
   \hline
    \hline
 NS-NS  & (n),(o) & $10^{10}$ | $\sim$ a few hundred years\\  
 BH-BH &(r), (s) & $\sim$ million years | $\sim$ hours\\
   \hline
	\end{tabular}
 \caption{Relevant GW merger time with SMBH, following one supernova and undergoing Hills process.}\label{table3}
\end{table}

We found that a wider initial condition for $a_1$  dramatically decrease the percentage of survived binaries systems with stellar (SMBH) companion. We quantify the Roche-limit crossing between the two inner members as  $a_{1,n}(1-e_{1,n})\leq R_{\rm Roche}$, and 
\begin{equation}\label{eq:Roche}
R_{\rm Roche} \sim 1.6 R_{2} \left(\frac{m_{2,n}}{m_{1}+m_{2,n}} \right)^{-1/3} \ ,
\end{equation}
where $R_{2}$ is the radius associated with $m_{2,n}$ which in this case is the radius of a Neutron star. Given these new orbits, for the wider initial conditions, we calculate the GW-emission time scale for the systems that survived SNe kicks and the systems that survived the kicks but the inner binary crossed the Roche limit of the tertiary. We found a shorter merger rates compared to Figure \ref{fig:nsbh_gwt} (not shown to avoid clutter). 

Our proof-of-concept simulations suggest that many of the survived systems will merge in less the the age of the Universe, with typical merger timescale of a few million years which can be as low as a few hundred years detectable by LISA (see Figure \ref{fig:LISA_prediction}). We consider the GW characteristic frequency ($f$) of the signal to be $f = v_p/r_p$, where $v_p$ and $r_p$ are the orbital velocity and the peri-center, and respectively 
\begin{equation}\label{eq:f}
    f =2\pi \frac{(1+e_{\rm Hill})^{1/2}}{(1- e_{\rm Hill})^{-3/2}}\frac{1}{P_{2,n}} \ ,
\end{equation}
where for simplicity we take the compact member (or the heavy member)  of the inner binary to be the one who will merge with the SMBH (denote $m$), and set the new semi-major axis to simply be $a_{2,n}$, and thus the $P_{2,n}$ is the post SN outer orbital period. The new eccentricity around the SMBH via the Hills process is estimates as: $e_{\rm Hill}\sim 1-(m/m_3)^{1/3}$ \citep[e.g.,][]{Hills88,Yu+03}. See table \ref{table3} for a range of merger times after the supernova.  These results suggest that a supernova remnant might serve as an signature for the process. 

In Figure \ref{fig:a2Cross6}, the ratio of $V_{3,n}/V_{c,esc}$ is color coded, where $V_{c,esc}=\sqrt{2\mu_3/a_{2,n}}$ is the post SN escape velocity from a circular orbit. As expected, further away from the SMBH the binary velocity is larger\footnote{ Note, that an interesting observational detection of    hyper-velocity late-type-B stars is consistent with the velocity associated with SN kick in the Galactic Center \citep[e.g.,][]{Tauris1998,Tauris2014,Tauris15}.  }. Note that systems that actually escape the SMBH potential well are not shown here, because they do not have a defined $a_{2,n}$.
We do find that some fraction of the binary systems (depending  on the initial conditions) will escape and unbind from the SMBH.  Run (j) (of uniform Monte-Carlo of the inner orbit's SMA, and extreme cusp distribution around the SMBH) suggests that about \textbf{$3$\%} out of all $29\%$ escaped binaries will be observed as hyper velocity binaries.

However, we note that those binaries with new velocities that are larger than $V_{c,esc}$, may still be bound to the galactic center by the potential of the bulge, disk or halo. To quantify this, we adopt a minimal velocity of $V_{3n}>200$~km~s$^{-1}$, following \citet{PortegiesZwart2006} who suggested that such a fast binary may escape the galactic center. For the NS-BH system discussed in this section, all of the systems that have $V_{3,n}>V_{c,esc}$, also have $V_{3n}>200$~km~s$^{-1}$. However, this behavior is highly sensitive to the initial conditions of the distribution of the binaries around the SMBH. In some later cases we find that more than half of the system with velocity larger than the escape velocity will still remain bound to the galactic center. Note that in Table \ref{table} and Table \ref{table E} we consider escape from the SMBH, since this proof-of-concept calculation constraint to the three-body regime. We further found that for BH-NS binary with stellar companion, a third of all the escaped binary systems would have $V_{3n}>200$~km~s$^{-1}$.

\begin{figure}[t!]
\begin{center}
\includegraphics[width=\linewidth]{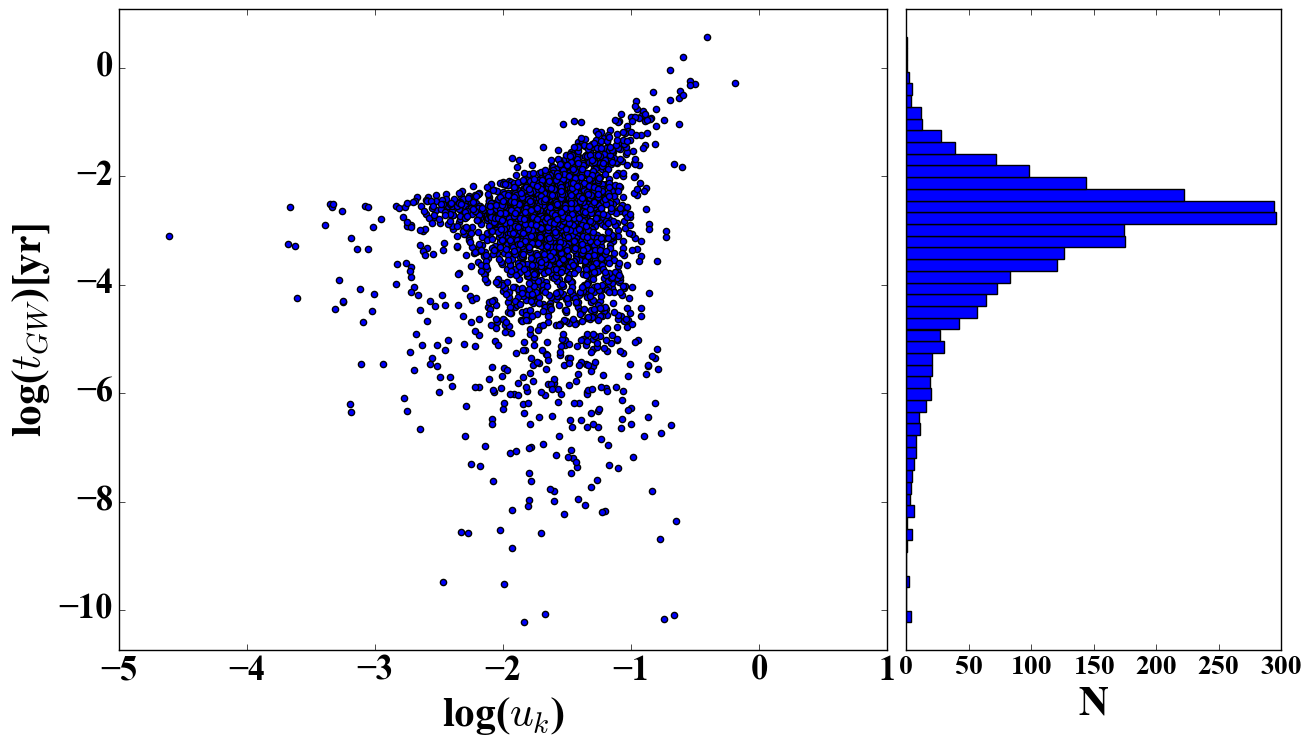}
\caption{\upshape BH-BH GW timescale after two SNs in systems around SMBH third companion with $a_1=5 R_\odot$ and $a_2$ from MC$^{\text{2,BW}}$, i.e., run (r). {\bf{Left Panel}}: We show the GW timescale as a function of the dimensionless kick velocity,  $u_k$. In the  {\bf{right Panel}} we show the histogram of the  GW timescales.  \vspace{0.3cm} }\label{fig:BH_gw}
\end{center}
\end{figure}

\begin{figure}[t!]
\begin{center}
\includegraphics[width=\linewidth]{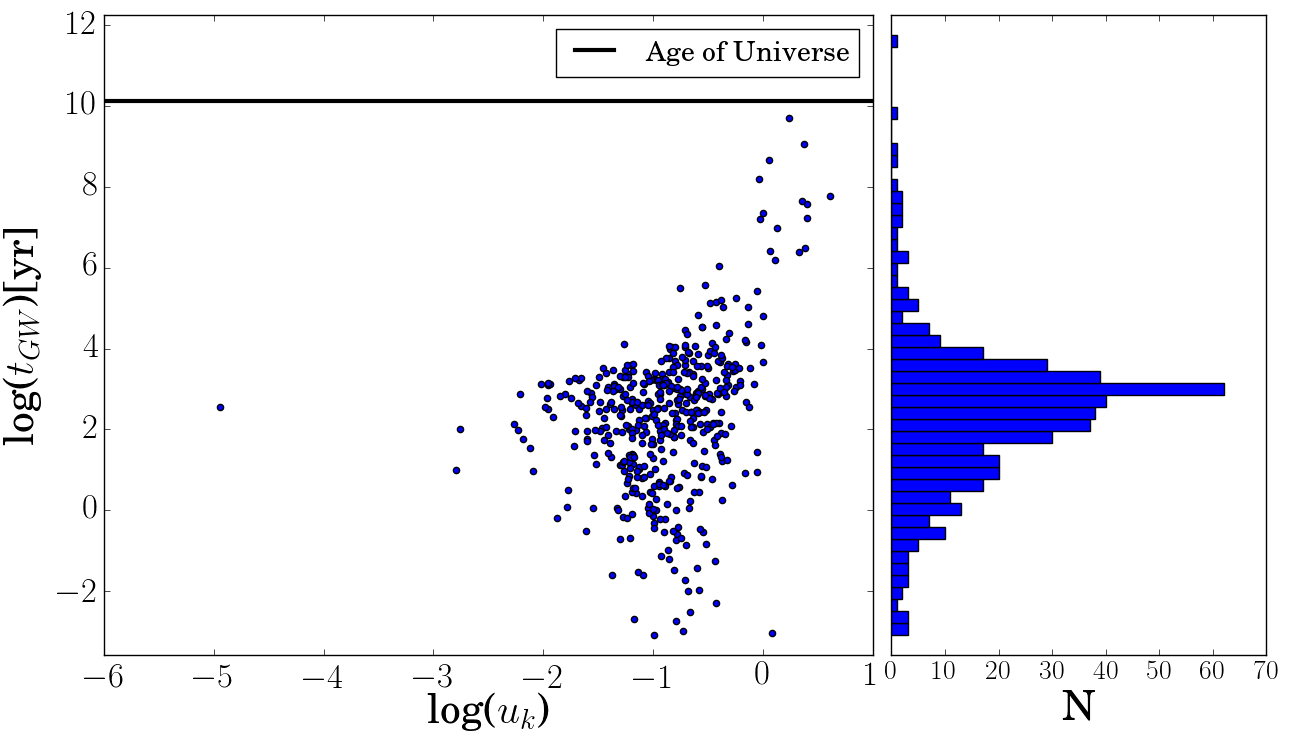}
\caption{\upshape NS-NS GW timescale after two SNs in systems around SMBH third companion with $a_1=5 R_\odot$ and $a_2$ from MC$^{\text{2,BW}}$, i.e., run (n).. {\bf{Left Panel}}: We show the GW timescale as a function of the dimensionless kick velocity,  $u_k$. In the  {\bf
{right Panel}} we show the histogram of the  GW timescales.  \vspace{0.3cm} }\label{fig:NS_gw}
\end{center}
\end{figure}

%% file: chapters/GW_source/bhb.tex
\subsubsection{ Black Hole binaries -  two natal kicks}\label{sec:LIGO-2BH}

The recent detections of Black Hole binaries (BHB) \citep{LIGO_PRL,LIGO_BH2,LIGO3} via LIGO revolutionized the field with the realization that the merger rate of BHB is now constrained to be between  $9- 240$~Gpc$^{-3}$~yr$^{-1}$ \citep{LIGO_BHm}. The astrophysical origin of these binaries is still under debate.

Many observational campaigns suggest that SN of BH progenitors  have no natal kick associated with them \citep[e.g.,][]{Willems+05,Reid+14,Ertl+16,Mandel16,Sukhbold+16}. However, \citet{Repetto+12} and \citet{Repetto+15} suggested that BHs likely receive natal kicks similar in magnitude to neutron stars\footnote{Although \citet{Mandel16} suggested that these studies overestimate the inferred natal BH kick distribution}. This is supported by the detection of one example of a non-negligible natal kick, \citep[e.g.,][]{Gualandris+05,Fragos+09}.  While it is still unclear what kick velocity magnitude if at all BH exhibit, it is clear that natal kick will affect  the orbital configurations of these systems.

As a proof-of-concept we adopt the aforementioned natal kick distribution for each BH member.  
We adopt the following parameters for a Monte-Carlo simulations: The inner BH-BH progenitor orbit SMA was chosen to be $a_{1}=5$~R$_{\odot}$ while the outer orbit SMA was chosen to be $a_{2}$=1000~AU. We also performed a Monte-Carlo simulation where $a_{2}$ was chosen from a uniform in log distribution (keeping $a_{1}=5$~R$_{\odot}$). The minimal value satisfied $\epsilon=0.1$ and a maximum value of $10,000$~AU. We chose the initial mass of the BH progenitors to be $m_1=15$~M$_\odot$ and $m_2= 31$~M$_\odot$, {\it just before the explosion\footnote{This means, of course, that the main sequence mass were much larger. }}.
%, taken from the stellar evolution code by \citet{Hurley+00}. 
The heavier star after mass loss went through mass loss due to SN explosion and reduced its mass to $m_{2,n}=30$~M$_{\odot}$. Shortly after the first SN event, the second star went through SN and results in $m_{1,n}=14$~M$_{\odot}$ BH \citep[e.g.,][]{Hurley+00}.   As before, for the tertiary body, we used $3$~M$_{\odot}$ for stellar companion and  $4\times10^6$~M$_{\odot}$ for SMBH companion. In the case of SMBH companion we also run a Monte-Carlo simulations choosing $a_2$ from a  Bahcall-Wolf-like distribution setting the density proportional to $a_2^{-2}$ \citep[e.g.,][]{Bahcall+76}, between $a_2=100$~AU and $a_2=0.1$~pc.  The other orbital parameters were chosen as explained above (see the beginning of \S \ref{sec:App}).

\begin{figure*}%[h!]
\begin{center}
\includegraphics[width=\linewidth]{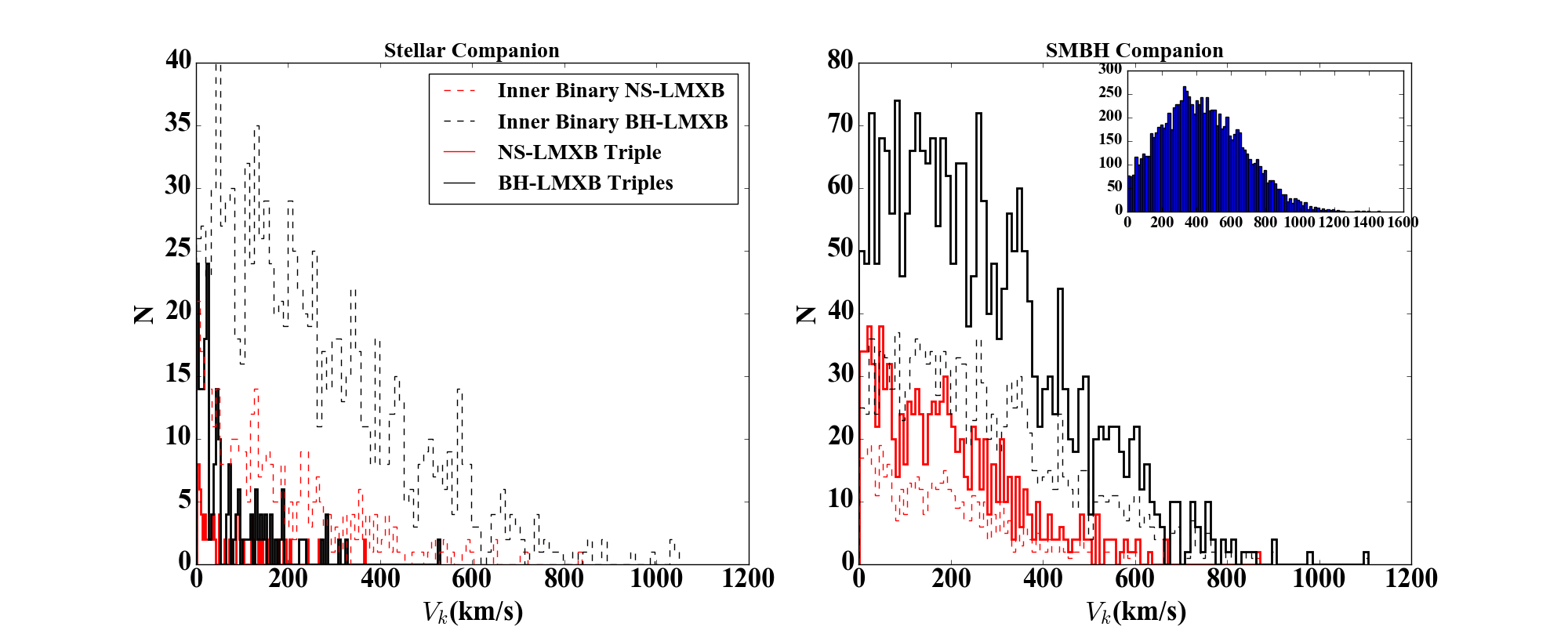}\label{fig:LMXBsur}%{Survived_1sn}old pic
\caption{\upshape Kick Velocity Distribution of Survived System in BH-LMXB (black lines), NS-LMXB (red lines), with an SMBH companion. This is for systems with $a_1$ chosen from MC$^{1}$ and $a_2$ from MC$^{\text{2,EC}}$. In the {\bf{left panel}} we consider a stellar companion as the tertiary $3$~M$_\odot$ while in the {\bf{right panel}} we consider $4\times10^6$~M$_\odot$ companion. We show the survived binary distribution for the systems in dashed lines while the solid lines depict the distribution of the survived triple systems.   The initial distribution (identical in both bases) is depicted in the inset.
%The survived systems with a distant low mass star companion after the SN. {\bf{ Right Panel}}: The survived systems with a distant Supermassive Blackhole companion after the SN. The solid lines shows the survived triple systems. The dashed lines represents the survived inner binaries and note that their survival rate is independent of the mass of the outer body. We conclude that with the SMBH companion, the surviving triples after SN saturates and equal to the number of surviving binaries. 
\vspace{0.3cm} }\label{fig:survived_1sn}
\end{center}
\end{figure*}

Figure \ref{fig:survived_2sn} depicts the retention numbers of BH-BH binaries and triples after each of the SN took place, (BH-BH are represented in black lines). As shown, large kicks yields larger escape velocities, which result in unbinding the binary and triple systems. The fraction of bound binaries after the first SN explosion is of course independent of the mass of the tertiary. However, the fraction of triples that remain bound after the first SN is different. As expected, the triple configuration is more likely to survive with a SMBH as the third companion after the first SN. Specifically, for BH-BH with fixed outer SMA at $1000$~AU, only  $15$ out of $10,000$ triple systems remain bound with a stellar tertiary and $4665$ with SMBH. For BH-BH binary system, upon 1st SN, near 50\% percent of binaries escape from its stellar companion. When the second SN took place about $0\%$ and $32\%$ of the systems remain bound in the triple configuration with a stellar and SMBH companion, respectively. About $98\%$ of BH-BH binary systems stay bound in a triple system with a SMBH companion, following two SN-kicks as large as the ones expected to take place in NS. 
%About \textbf{$0\%$} of the systems with stellar companion (run MC$^2$)  had one of the BHs that crossed the outer orbit's Roche radius, which may result in BH-LMXB.
The kicks expected to take place in BH-BH systems are typically much lower or not at all (see above discussion). Thus, this result suggests that the fraction of BH-BH binaries to exist near galactic nuclei is rather large, which later can merge via GW emission \citep[e.g.,][]{Hoang+17}.  We estimate the GW timescale at which these binaries will merge (see Figure \ref{fig:BH_gw}), in the absence of Eccentric Koza-Lidov mechanism, and find an average merging time of $\sim$ 1 million years and reduce to an extremely short timescale $\sim 10^{-2}$ years after 2SN and thus resulting in the merger of BH-SMBH detectable by LISA with an SN progenitor. 

As mentioned in Appendix \ref{App:5au}, we also considered a wider initial inner binary ($a_1=5$~au). These wider initial systems increase the  percentages of systems that cross the stellar (SMBH) companion Roche limit. For example we consider run (s) for $a_1=5$~R$_\odot$ and run (s5) for $a_1=5$~au, which represent a BH binary at $1000$~au around an SMBH. The tighter initial configuration had about $1\%$ of the system crossing the SMBH Roche limit while the wider resulted in $\sim 54$\% (see Tables \ref{table} and \ref{table E}).  Specifically we consider post the 1st SN (Figure \ref{fig:bigplot_1sn}) and after the 2nd SN (Figure \ref{fig:bigplot_2sn}). We also present the histogram post the two SNs  of these systems,  Figure \ref{fig:BH_grid}. 

%% file: chapters/GW_source/ns_ns.tex
\subsubsection{Double neutron star - two natal kicks}\label{sec:LIGO-2NS}

Recently LIGO detected a GW signal from a NS-NS merger  \citep{LIGO2NS} and its associated electromagnetic counterpart \citep[e.g.,][]{Abbott+17,Alexander+17,Blanchard+17,Chornock+17,Fong+17,Margutti+17,Nicholl+17,Soares+17,Coulter+17}. This detection was long expected \citep[e.g.,][]{Acernese+08,Abbott+09,Somiya+12} in light of the expected abundance of NS binaries.
There are currently about $70$ NS binary (or  Double Neutron Star) systems detected \citep[e.g.,][]{Lattimer+12,Ozel+16} out of  thousands of known NSs \citep[e.g.,][]{Hobbs+05,Manchester+05}.  Double neutron stars   are the  prime candidate progenitors for short Gamma-Ray burst events  \citep[e.g.,][]{Eichler+89,Metzger+15}, and are also the  main candidate for heavy r-process nucleosynthesis sources  \citep[e.g.,][]{Lattimer+74,Eichler+89,Beniamini+16,Beniamini+16Delay}. 
The recent detection of GW170817 GW from the coalescence of double NS combined with gamma-ray burst (GRB 170817A) with subsequent ultraviolet, optical, infrared observations  \citep{LIGO2NS} showed that these theoretical models are very promising.  
Furthermore, \citet{LIGO_NS} placed an upper limit on the merger rate to be $12,600$~Gpc$^{-3}$~yr$^{-1}$. NS binaries (for initially circular systems) have a substantial probability of getting disrupted when one of the stars goes through a SN, either because the instantaneous mass loss associated with the SN or because of the resulting asymmetry in the imprinted natal kick of the newborn NS. 

To explore kicks in NS-binary in a triple configuration, we adopt a nominal proof-of-concept system composed with $m_1=4$~M$_\odot$, which leaves a NS with $m_{1,n}=1.4$~M$_\odot$, and $m_2=5$~M$_\odot$, which leaves a NS with $m_{2,n}=1.4$~M$_\odot$, and $a_1=5$~R$_\odot$, $a_2=1000$~AU. As usual, we had two choices for $m_3$; in the first we set $m_3=3$~M$_\odot$ assuming a stellar companion and in the second we assume that the NS-NS binary is located in the galactic center and set $m_3=4\times 10^6$~M$_\odot$. 

For this numerical example (case (n)) we adopt {\it two} SN natal kicks with each NS. Each SN-kick is adopted with the same normal distribution described in \S \ref{sec:LIGO-NS-BH}. {As expected the second SN-kick significantly reduces the fraction of survived binaries and triples as depicted with red lines in the right panel of Figure 
\ref{fig:survived_2sn} compared to the left panel (1SN), as we see that both dotted and solid red curves have smaller overall amplitudes.}  In fact we find that non of our proof-of-concept NS-NS binaries with {\it stellar} companion remained in triple configuration (as of case(l) and (m)). That is not surprising as we choose a less massive tertiary ($m_3=3$~M$_\odot$). Again, the triple configuration is more likely to remain bound in the presence of a massive tertiary. {Compared to the stellar tertiary which is not able to keep any triple systems bound, we see $20$\% of triples remains bound out of binaries with a SMBH companion(e.g. see comparison between case (i) and (o) in Table \ref{table}).} An SMBH companion will keep a large fraction of the binaries remaining in their triple configuration, which is somewhat sensitive to the outer orbit initial separation, see Table \ref{table}. In our test of Bahcall-Wolf distribution (case (n)) we found that about $57\%$ of all survived binaries escaped the system.  Thus according to our definition, $ 6\%$ these systems will become hyper velocity binaries. We also note that a change in the initial tight inner binary SMA $a_1$ from $5 R_{\odot}$ to $5$AU would dramatically change the percentage of survived binaries systems with stellar(SMBH) companion crossed the Roche limit of tertiary BH. In the most dramatic case, the percentage change from $3$\% to $\sim 84$\% (see Table \ref{table E} case (o) and (o5) post 1st SN).

We also calculate the merger time via GW emission of the two NS following the kicks and find that for these proof-of-concept initial conditions. {After the first SN, the GW merger time is on average longer than the age of the Universe. However, after the 2nd SN, the merging time is relatively short, on average takes $200$~years to merge via GW, as depicted in Figure \ref{fig:NS_gw}}

We note that  \citet{Michaely+18} found that in some cases the SN-kick caused the NS-NS systems to reach such small separations that they cross the Roche limit or even immediately merge via GW emission. Since in our numerical experiments we have made sure to place the tertiary far away such that $\epsilon<0.1$, on average the GW emission will have about $200$~yr delay, while a small fraction of them will have only few days (and as low as few hours) delay (see  Figure \ref{fig:NS_gw}).

%% file: chapters/Xray_binary/ns_lmxb.tex
\subsection{Low mass X-ray Binaries}\label{sec:LMXB}

A substantial number of close binaries with an accreting 
compact object, mainly Low mass X-ray Binaries (LMXBs) and their descendants (i.e., millisecond radio pulsars) are known or suspected triples  
\citep{T93,T99,rasio01,sig03,Chou,1820,Corbet94,Grindlay88,BG87,Garcia,Prodan:2012ey}.Furthermore, it was recently suggested that the inner 1pc of the Galactic center host an over abundance of X-ray binaries \citep[e.g. see][]{Hailey2008} Thus, a natural question is what is the probability that these systems will remain in their triple configuration after the SN natal kick took place. 

In each case of BH- and NS- LMXB we have three Monte-Carlo tests. In all we choose $a_1$ from a Monte-Carlo simulation labeled $MC^1$ which is uniform in log between $5$~R$_\odot$ and a $1000$~R$_\odot$. We also choose $a_2$ to follow either MC$^{2,EC}$ (consistent with $a_2^{-3}$), or MC$^{2,BW}$ \citep[consistent with $a_2^{-2}$, e.g.,][]{Bahcall+76}. As before we have two masses for the third companion, stellar companion with $3$~M$_\odot$ and a SMBH with $4\times 10^6$~M$_\odot$. See Table 1 for more information. We chose the inner and outer orbital eccentricity from a uniform distribution \citep[e.g.,][]{Raghavan+10}, and the mutual inclination is chosen from an isotropic distribution (i.e., uniform in $\cos i$). In addition the inner and outer argument of perihapsis angles are chosen from a uniform distribution between $0-2\pi$.

\subsubsection{Neutron Star Low mass X-ray Binaries}\label{sec:LMXB-NS}

In this case, we adopt a nominal system composed of $m_1=1$~M$_\odot$, $m_2= 4$~M$_\odot$, which leaves a NS with $m_{2,n}=1.4$~M$_\odot$.  
We have tested three simplified Monte-Carlo runs (see Table \ref{table}), and found that only $4\%$ of all binaries remain bound after the kick, for both SMBH and stellar third companion. 
Considering the stellar tertiary proof-of-concept test,  i.e., run (a), we found that the  SN-kicks disrupt all of the stellar triples. Considering a SMBH triple companion (runs  (b) and (c)), we find that $94\%$ and $99\%$, respectively, out of the surviving binaries will stay bound to the SMBH companion (see Figure \ref{fig:survived_1sn}). The remaining binaries will escape the SMBH potential well at velocity smaller than $200$ km s$^{-1}$.

In addition, we found that $19\%$ of all survived binary systems cross the low mass star Roche limit and start accreting onto the NS. In other words  $a_{1,n}(1-e_{1,n})\leq R_{\rm Roche}$, and 
\begin{equation}\label{eq:Roche}
R_{\rm Roche} = 1.6 R_{\star} \left(\frac{m_{\star}}{m_{NS}+m_{\star}} \right)^{-1/3} \ ,
\end{equation}
where $R_{\star}$ is the radius of the star. These systems are forming NS-LMXB immediately after the SN. 

We also found that $4\%$ out of the triples systems crossed the tertiary SMBH companion's Roche limit according to Equation (\ref{eq:a2Cross}). In the case of SMBH tertiary these NSs will merge with the SMBH by emitting gravitational waves after about a million years on average. Thus, producing a GW-LISA event with a possible young SN remnant. 

%% file: chapters/Xray_binary/bh_lmxb.tex
\subsubsection{ Black Hole Low mass X-ray Binaries}\label{sec:LMXB-BH}

%\subsection{Black Hole Low mass X-ray Binaries}
The formation of BH-LMXB poses a theoretical challenge, as low-mass companions are not expected to survive the common-envelope scenario with the BH progenitor \citep[see][]{Podsiadlowski+03}. Recently, \citet{Naoz+16} proposed a new formation mechanism that skips the common-envelope scenario and relies on triple-body dynamics. Specifically, using the eccentric Kozai-Lidov mechanism \citep[e.g.,][]{Naoz16} they showed that eccentricity excitations due to gravitational perturbations from a third star can rather efficiently form BH-LMXB. Their calculations assume no SN-kicks, consistent with observational and theoretical studies \citep[e.g.,][]{Willems+05,Reid+14,Ertl+16,Mandel16,Sukhbold+16}. However, at least in one system robust evidence for a non-negligible natal kick imparted onto a BH system was detected \citep[e.g.,][]{Gualandris+05,Fragos+09}.  Here we show that even given large SN-kicks in these systems, it still allows for large fraction of these systems to remain bound.  

 We adopt a nominal system of $m_1=1$~M$_\odot$, $m_{2}=9$~M$_\odot$, which leaves a BH with a mass of $m_{2,n}=7$~M$_\odot$ \citep[which follows the stellar evolution adopted from {\tt SSE} code][]{Hurley+00}.  We chose our orbital parameters as explained above. The kick magnitude distribution was chosen in the same way as described above (see \S \ref{sec:LMXB-NS}). We find that about $11\% $ of the binaries survived (as depicted in Figure \ref{fig:survived_1sn}).  For BH-LMXB with stellar mass companion, we found 11\% of binaries (out of all 99\%) escape from its stellar companion at hyper velocity.Furthermore, we find that about $1\%$ of the surviving binaries remained in their triple configurations for stellar mass companions and as expected $99\%$ for the SMBH companion. 
  Not surprisingly, this is a larger fraction than the  surviving NS-LMXB triple system, as (i) the BH mass has a larger gravitational potential, and (ii) the mass loss was substantially a smaller fraction of the initial mass compared to the NS explosion. Thus, we find that \citet{Naoz+16} BH-LMXB mechanism can still work even in the presence of large natal SN-kicks for the BH.

%We use this proof-of-concept numerical experiment to test one of our analytical equations. Specifically,   Equation (\ref{eq:e1n}) yields a simple relation between the pre- and post-SN eccentricity of the inner binary. We  show this analytical expression for the maximum $\bf u_k$ (black lines) in Figure \ref{fig:LMBXe1}, where the Monte-Carlo surviving systems occupy the part of the parameter space between this line.   

 In our simulations we found that $\sim 24\%$ of all binaries cross their Roche limits (see Eq.~(\ref{eq:Roche})) and thus form BH-LMXB. We also found that $\sim 13\%$ of all triples cross the stellar third companion's Roche limits according to Equation (\ref{eq:a2Cross}), again forming BH-LMXB, immediately after the SN. 
 
 In the case of binaries around SMBH, we found that in $7\%$ of triple systems, the newly formed stellar mass BH crosses the SMBH Roche limit, thus merging with the SMBH on a typical timescale of {$10$ million years} and can be as short as few minutes (see Figure \ref{fig:LISA_prediction}). This potentially forms a systems detectable by LISA after the SN, thus allowing for an optical precursor counterpart appearing shortly before the GW detection.

 %n our simulations, a fraction of our NS-LXMB binary will collide or merge shortly after the SN. With stellar companion, 23.52\% of the inner binary will get so close that they will cross Roche limit of their companion. With supermassive black hole as a companion, the percentage of the colliding inner binary decreases and 4.9\% of of the binary will collide within an orbital period.

%% file: chapters/discussion.tex
\section{Discussion}\label{sec:dis}

We analyzed SN-kicks in triple configurations. In recent years, hierarchical triple body have been proven to be very useful in addressing and understanding the dynamics of various systems from exoplanets to triple stars and compact object systems \citep[see][and references therein]{Naoz16}. As a star undergoes SN and forms a NS (or BH) it is expected to have a natal kick. In a binary system this kick may cause the velocity vector orientation  and amplitude, of the mutual center of mass,  to vary. The consequences of such a kick in a {\it binary} system has been previously investigated in the  literature, often focused on circular orbits. With the gaining interest in triple systems, we address the natal kick consequences in the context of triple systems with eccentric orbits. 

We have derived the analytical equations that describe the effect of a natal kick in hierarchical triple body systems. Triple systems have been considered in the literature as a promising mechanism to induce compact object binaries through GW emissions \citep[e.g.,][]{Blaes2002,Tho10,Antonini+10, Pijloo2012, Michaely+16, silsbee17,VanLandingham16, Hoang+17, Petrovich+2017, Hamers+18CO,Randall2018a,Randall2018b}. We consider the effects of SN-kick in keeping compact object binaries and triples on bound orbits. Furthermore, we also consider the effects of kicks in producing LMXB. 

We have run proof-of-concept Monte-Carlo simulations to test the range of applications of SN-kicks in hierarchical triples. Below we summarize our significant results.

\begin{itemize}
    \item {\bf SN-kicks may shrink the binary SMA.} We pointed out that SN-kicks can lead the  binary SMA to {\it shrink} in many cases, as can be seen from Equation  (\ref{eq:a2n}). See for example Figure \ref{fig:NSBH_a1} for the agreement between the analytical and numerical results for the shrinking condition.  In fact, we found that a combination of expanding of the inner orbit and shirking of the outer orbit cause a non-negligible fraction of our systems to cross the inner binary's Roche limit as well as the outer orbit's Roche limit, resulting in destruction of the triple system and a possible merger with the tertiary companion. Shrinking the inner orbit binary SMA can trigger a merger or a common envelope event. Crossing the tertiary's Roche limit has far more dramatic consequences as we elaborate below. 
    
    \item {\bf Massive tertiaries}. As expected, we find  that  the triple configuration remains bound when the tertiary is more massive (see Figure  \ref{fig:mass}, and compare the right and left panels in Figures \ref{fig:survived_nsbh}, \ref{fig:survived_2sn} and \ref{fig:survived_1sn}). This trend has significant implications on the formation of LMXBs as well as GW emission from compact objects.  
    
    \citet{Naoz+16} recently suggested that gravitational perturbations from a distant companion can facilitate the formation of LMXBs. Thus, overcoming the nominal theoretical challenge associated with the BH-LMXB formation, since low-mass companions are not expected to survive the common-envelope phase with the BH progenitor. Therefore, our result that the majority of binaries remain near a SMBH post-SN kick (e.g., Figure \ref{fig:survived_1sn}), suggests that SMBH environment yield a larger abundance of LMXBs. 
    
    Furthermore, the SMBHs gravitational perturbations can enhance the compact objects merger rate \citep[e.g.,][]{Antonini+14}, which can result in a non-negligible rate from $1-14$~ Gpc$^{-3}$~yr$^{-1}$ \citep{Hoang+17}. Our results here suggest that the majority of binaries  that survive the SN-kick will not escape the SMBH potential wells, as expected.

    \item {\bf Hyper velocity binaries}. When the tertiary companion is a star, the SN-kick tends to simply disrupt triple stellar system. However, if the  tertiary is a SMBH a SN-kick leads to a few percent of the binaries escaping the SMBH potential well and will be observed as hyper velocity binary system.
    
    \item{\bf Simultaneous and precursors electromagnetic signatures for LIGO compact object merger event.} We find that since the SN-kick can shrink the SMA of the binary orbit, it leads to a short GW emission merging time, which will be prompted by SN. This type of behavior was pointed out previously by \citet{Michaely+18} for inner binaries. We find similar results, for our NS-BH, BH-BH and NS-NS proof-of-concept examples (e.g., Figures \ref{fig:nsbh_gwt}, \ref{fig:BH_gw} and \ref{fig:NS_gw}). Consistently with  \citet{Michaely+18}, we find that many of these LIGO events will have a supernova remnant signature. Furthermore, for the systems that underwent two SN-kicks we find that the SN can be either as a    precursors or even almost simultaneous with the GW detection. For example, as shown in Figure \ref{fig:BH_gw}, for BH-BH merger, on average, the SN precursor takes place about a month before the GW detection, with some that will take place almost instantaneously.  
    
    \item {\bf Electromagnetic precursors for LISA events.} Finally we find that the SN-kick causes a non-negligible fraction of the systems near an SMBH to cross the SMBH Roche limit. If the companion star crosses the Roche limit of SMBH, it will cause a tidal disruption event {\bf TDE}, shortly after the SN. Interestingly, if the compact object crossed the SMBH Roche limit, {\bf  resulting an Extreme mass ratio inspiral (EMRI). }Thus, we find that GW emission might result in a LISA event after a wide range of times, which depends on the Monte-Carlo configuration. On average, events in the LISA detection band  will take place about {a few million} years after the Supernova, and they can be as quick as a few minutes after the supernova explosion. 
   
   \end{itemize}

 We have tested a wide range of initial conditions, from a tight binary ($a_1=5$~R$_\odot$, which represents most of the discussion throughout the paper) to a wide initial binary ($a_1=5$~au, see Appendix \ref{App:5au}). The former may represent a stable binary that survived a common-envelope evolution. In both cases, the qualitative result seems to hold, but the fraction, as expected differs. For example, the fraction of binaries that survive SN kick diminishes, but the fraction of systems that crossed the tertiary Roche-limit gone up. The latter is especially interesting, as it can cause either a TDE or an EMRI with a possible supernova precursor.